\documentclass[fleqn,usenatbib]{mnras}
\usepackage{newtxtext,newtxmath}
\usepackage[T1]{fontenc}
\DeclareRobustCommand{\VAN}[3]{#2}
\let\VANthebibliography\thebibliography
\def\thebibliography{\DeclareRobustCommand{\VAN}[3]{##3}\VANthebibliography}

\usepackage{graphicx}	%
\usepackage{amsmath}	%

\usepackage{hyperref}
\usepackage{fixme}
\fxsetup{
  status=draft,
  author=CC,
  layout=inline,nomargin,nofootnote,
  theme=colorsig,
}

\FXRegisterAuthor{cc}{acc}{CC}
\FXRegisterAuthor{mrk}{amrk}{MRK}

\newcommand{\quokka}[0]{\textsc{Quokka}}
\newcommand{\betasq}{\frac{v^2}{c^2}}

\newcommand{\vecx}{\boldsymbol{x}}
\newcommand{\vecv}{\boldsymbol{v}}

\newcommand{\vecG}{\boldsymbol{G}}
\newcommand{\vecF}{\boldsymbol{F}}
\newcommand{\vecy}{\boldsymbol{y}}
\newcommand{\vecs}{\boldsymbol{s}}
\newcommand{\vecf}{\boldsymbol{f}}
\newcommand{\tenP}{\mathsf{P}}

\newcommand{\tenI}{\mathsf{I}}
\newcommand{\hatt}{\hat{t}}
\newcommand{\hrho}{\hat{\rho}}

\newcommand{\hE}{\hat{E}}
\newcommand{\hnabla}{\hat{\nabla}}
\newcommand{\hp}{\hat{p}}
\newcommand{\hv}{\hat{v}}

\newcommand{\hF}{\hat{F}}
\newcommand{\hx}{\hat{x}}
\newcommand{\hB}{\hat{B}}
\newcommand{\hP}{\hat{P}}
\newcommand{\hT}{\hat{T}}
\newcommand{\hG}{\hat{G}}

\newcommand{\hchi}{\hat{\chi}}
\newcommand{\hvecv}{\hat{\boldsymbol{v}}}

\newcommand{\hvecF}{\hat{\boldsymbol{F}}}
\newcommand{\hvecG}{\hat{\boldsymbol{G}}}
\newcommand{\htenP}{\hat{\mathsf{P}}}
\newcommand{\hvecFU}{\hat{\boldsymbol{F}}_{\hat{\boldsymbol{U}}}}
\newcommand{\hvecU}{{\hat{\boldsymbol{U}}}}

\newcommand{\hvecS}{\hat{\boldsymbol{S}}}
\newcommand{\calC}{\mathcal{C}}
\newcommand{\calP}{\mathcal{P}}

\newcommand{\calL}{\mathcal{L}}

\newcommand{\chat}{\hat{c}}

\newcommand{\aref}[1]{\hyperref[#1]{Appendix~\ref{#1}}}

\title[Asymptotic-preserving Implicit-explicit Method]{An Asymptotically-Correct Implicit-Explicit Time Integration Scheme for Finite Volume Radiation-Hydrodynamics}

\author[He, Wibking, \& Krumholz]{
Chong-Chong He$^{1}$\thanks{E-mail: Chongchong.He@anu.edu.au (CCH)}, Benjamin D. Wibking$^{2}$, and Mark R. Krumholz$^{1,3}$
\\
$^{1}$Research School of Astronomy and Astrophysics, Australian National University, Canberra, ACT 2611, Australia\\
$^{2}$Department of Physics and Astronomy, Michigan State University, East Lansing, MI 48824, USA\\
$^{3}$ARC Centre of Excellence for Astronomy in Three Dimensions (ASTRO-3D), Canberra, ACT 2611, Australia
}

\date{Accepted XXX. Received YYY; in original form ZZZ}
\pubyear{2022}

\begin{document}
\label{firstpage}
\pagerange{\pageref{firstpage}--\pageref{lastpage}}
\maketitle

\begin{abstract}
Numerical radiation-hydrodynamics (RHD) for non-relativistic flows is a challenging problem because it encompasses processes acting over a very broad range of timescales, and where the relative importance of these processes often varies by orders of magnitude across the computational domain. Here we present a new implicit-explicit (IMEX) method for numerical RHD that has a number of desirable properties that have not previously been combined in a single method. Our scheme is based on moments and allows machine-precision conservation of energy and momentum, making it highly suitable for adaptive mesh refinement applications; it requires no more communication than hydrodynamics and includes no non-local iterative steps, making it highly suitable for massively parallel and GPU-based systems where communication is a bottleneck; and we show that it is asymptotically-accurate in the streaming, static diffusion, and dynamic diffusion limits, including in the so-called asymptotic diffusion regime where the computational grid does not resolve the photon mean free path. We implement our method in the GPU-accelerated RHD code \quokka{} and show that it passes a wide range of numerical tests. 
\end{abstract}


\section{Introduction}
\label{sec:intro}

Radiation-hydrodynamics (RHD) plays a significant role in astrophysics, influencing the evolution and energy distribution in various astrophysical systems or phenomena -- stellar atmospheres \citep[e.g.][]{Mihalas1978}, planetary atmospheres \citep[e.g.][]{Zhang2020}, core-collapse supernovae \citep[e.g.][]{Skinner2016, Radice2018}, star formation in a variety of environments \citep[e.g.][]{Thompson2005, Krumholz09c, Rosen16a, He2019, Menon22c}, active galactic nuclei and jets \citep[e.g.][]{Davis2020}, and galactic outflows \citep[e.g.][]{Naab2017, Zhang2020}. 
The dynamics of RHD systems vary substantially in a range of scales and physical conditions, parametrized by the typical optical depth and the sound speed of radiation-interacting gas, which determines how radiation is transported. 
While several numerical techniques exist to solve the RHD equations in various limits, developing methods that are accurate across all regimes and that run efficiently on modern, GPU-based architectures remains an ongoing challenge. 

There are well-known difficulties associated with solving the RHD equations numerically for non-relativistic systems. One significant challenge arises from the huge difference in timescales associated with radiation and hydrodynamics -- characterised by the speed of light $c$ and gas flow speed $v$, respectively. A second is that realistic RHD problems often contain a huge range of opacities, such that the photon mean free path may be comparable to the size of the entire simulation domain in some regions, while in others it may be so small as to be impossible to resolve at reasonable computational cost. The RHD equations are well-suited to hydrodynamics-like explicit solution methods in some regimes, but often the source terms coupling the gas and radiation are so stiff that an implicit method must be used to ensure stability. 

To deal with these challenges, it is natural to solve both the transport and source terms in the RT equations implicitly so that the time step is not limited by the speed of light. This is the approach adopted by many RHD codes \citep[e.g.][]{Krumholz2007,Zhang11a,Jiang12a, Menon2022}. However, this approach suffers from the need for an implicit treatment of the transport term, which is non-local. In multiple dimensions, implicit update of this term is usually accomplished by solving a sparse matrix system, and sparse matrix solvers show limited scalability and performance on modern massively parallel and GPU-accelerated architectures, where the high and unpredictable communication load they involve becomes a performance bottleneck. 

Several authors have explored alternative approaches in which the source and transport terms are operator-split, with the former treated explicitly and the latter implicitly \citep[e.g.,][]{Rosdahl2015, Skinner2019, Wibking2022}. These schemes generally achieve better scaling and speed, particularly when running in parallel on large numbers of GPUs. One of the disadvantages of this approach is that the explicit treatment of the transport term requires a time step significantly smaller than the hydrodynamics one (since $c \gg v$), but it is possible to alleviate this problem by sub-cycling the radiation transport step relative to hydrodynamics. Since the transport equations for radiation are much simpler than those for hydrodynamics, it is possible to carry out many radiation transport updates per hydrodynamic update at a comparable cost. If necessary the cost can be mitigated further using the reduced speed of light (RSLA) approximation \citep[RSLA;][]{Gnedin01a, Skinner13a}. 

A second challenge in the operator-split approach has received less attention in the literature, but is perhaps even more serious: the need to properly balance the transport and source terms across all asymptotic regimes. This balance is critical because the RT equation behaves as an advection equation in optically thin regimes, but transitions to a diffusion equation in optically thick regimes, and near-perfect cancellations between parts of the source and transport terms are responsible for this behaviour. Efforts to recover this property of the RHD system have thus far mostly involved {\it ad hoc} corrections to the Riemann solver or to the source terms to recover the correct asymptotic limit. For instance, \cite{Rosdahl2015} add a ``trapped photons'' term to the source term to account for diffusion. \cite{Skinner2019} apply corrections to the wave speed in the Riemann solver, scaling down the characteristic speed of the radiation modes in optically thick regimes by a factor of $1/\sqrt{N \tau_{\rm cell}}$, where $\tau_{\rm cell}$ is the optical depth per cell, and $N$, chosen empirically, is a small integer. A similar approach is used by \cite{Wibking2022}. Despite these efforts, such corrections are often only partly successful, and their accuracy across a wide range of parameter space has not been tested. Perhaps the most successful (in terms of accuracy) operator-split method published to date is the discontinuous Galerkin implicit-explicit (DG-IMEX) scheme proposed by \citet{McClarren2008}, but as we discuss below even this scheme is not accurate in all RHD limits, and it has a number of other undesirable properties as well.

This situation motivates our goal of designing a method that achieves the best of both worlds: accuracy in all RHD regimes that is comparable to that achieved by fully-implicit methods, but without the need for poorly-scaling communication-intensive operations like sparse matrix solves. In this paper, we describe a method that achieves this goal using a novel time-integration scheme that recovers the proper asymptotic limit in the radiation diffusion regime without requiring non-local implicit updates. Our method is based on a convex-invariant, asymptotic-preserving IMEX approach that is second-order accurate in streaming limit, wherein the transport terms are handled explicitly, while the matter-radiation interacting part is treated implicitly and {\it locally}, eliminating the need for non-local implicit terms in iteration.

We begin in \autoref{sec:rhd} by introducing the full set of two-moment RHD equations to be solved and deriving characteristic numbers and limiting behaviours for them, laying the foundation for our analysis. Then, in \autoref{sec:imex}, we present the IMEX scheme. In \autoref{sec:asymp}, we derive some properties of our scheme and compare to alternative approaches. Finally, we present in \autoref{sec:test} tests of our implementation, demonstrating its effectiveness and applicability.

\section{The RHD equations and numerical methods for solving them} 
\label{sec:rhd}

We begin our analysis by describing the RHD system of equations in \autoref{ssec:rhd_system}, non-dimensionalising it to obtain characteristic numbers and limiting regimes in \autoref{ssec:characteristics}, and then using those to gain insight into the challenges of designing numerical methods for RHD and to motivate our approach in \autoref{ssec:rhd_methods}.

\subsection{The RHD system}
\label{ssec:rhd_system}

Our method begins from the fundamental equations of RHD\footnote{While in this paper we focus on RHD for simplicity, our method applies equally well to radiation-magnetohydroynamic systems and equations.} expressed in an inertial (lab) reference frame. These are
\begin{align} \label{eq:hyper}
  \frac{\partial \boldsymbol{U}}{\partial t}+\nabla \cdot\boldsymbol{F}_{\boldsymbol{U}}=\boldsymbol{S}_{\boldsymbol{U}},
\end{align}
where
\begin{equation}
  \label{eq:G}  
  \boldsymbol{U}=\left[
    \begin{array}{c}
      \rho \\
      \rho \boldsymbol{v} \\
      E_{\rm gas} \\
      E \\
      \frac{1}{c^2} \boldsymbol{F}
    \end{array}\right], \;
  \boldsymbol{F}_{\boldsymbol{U}} = \left[
    \begin{array}{c}
      \rho \boldsymbol{v} \\
      \rho \boldsymbol{v} \otimes \boldsymbol{v}+p \\
      (E_{\rm gas} + p) \boldsymbol{v} \\
      \boldsymbol{F} \\
      \tenP
    \end{array}\right], \;
  \boldsymbol{S}_{\boldsymbol{U}}=\left[
    \begin{array}{c}
      0 \\
      \boldsymbol{G} \\
      c G_0 \\
      - c G_0 \\
      - \boldsymbol{G} 
    \end{array}\right].
\end{equation}
are vectors of the conserved quantities, the advection terms, and the source terms, respectively; in the equations above, $\rho$ is the matter density, $\vecv$ is the matter velocity, $p$ is the matter pressure, $E_{\rm gas}$ is the gas total energy density, $E$ is the radiation energy density, $\vecF$ is the radiation flux, $\tenP$ is the radiation pressure tensor, and $(cG_0, \vecG)$ is the radiation four-force. A significant advantage of working with lab-frame rather than comoving-frame radiation quantities is that these equations are manifestly conservative, a feature that the algorithm we describe below will preserve. As usual in the moment formulation, however, one must adopt a closure relation for the radiation pressure tensor $\tenP$. There are a wide range of possible closures, and since our scheme is independent of this choice, we will not discuss closure relations further here.

Our next step is to write out the radiation four-force in the mixed-frame formulation, whereby we write  the matter-radiation exchange coefficients in the comoving frame, where they are simplest, while all other quantities remain in the lab frame. We assume that the emitting matter is in local thermodynamic equilibrium, so its emissivity is proportional to the Planck function, and we neglect scattering. To order $v^2/c^2$ the result, expressed in index notation and adopting the Einstein summation convention, is \citep{Krumholz2007} 
\begin{align}
  -c G_0 & =c \left(\chi_{0P} \frac{4 \pi B}{c} - \chi_{0E} E\right)\left(1+\frac{1}{2}\betasq\right) + (2\chi_{0E} - \chi_{0F}) \left(\frac{v_i F_i}{c} \right)
  \nonumber \\
  & {}
  + c(\chi_{0F} - \chi_{0E}) \left(\betasq E + \frac{v_i v_j P_{ij}}{c^2}\right), \label{eq:G0} \\
  -G_i & =-\chi_{0F} \frac{{F}_i}{c} \left(1+\frac{1}{2} \betasq\right) + \chi_{0P} \frac{4 \pi B}{c} \frac{v_i}{c} + \chi_{0F} \frac{{v_j} P_{ji}}{c}
  \nonumber \\
  & {}
  + (\chi_{0F} - \chi_{0E}) \left(E - \frac{2 v_j F_j}{c^2}\right) \frac{{v_i}}{c}.
  \label{eq:G1}
\end{align}
Here a subscript 0 in $\chi$ indicates the absorption coefficient is expressed in the comoving rather than the lab frame, and $\chi_{0P}$, $\chi_{0E}$, and $\chi_{0F}$ are the comoving-frame Planck, energy, and flux mean absorption coefficients. The leading-order part of the time-like component of the mixed-frame radiation four-force $G_0$ is the classical rate of radiation-matter energy exchange in the comoving frame, while the order $v/c$ part combines the work done by the radiation force on matter ($-\chi_{0F} v_i F_i/c$) with a purely relativistic effect arising from the transformation of the opacity between the comoving and lab frames ($2\chi_{0E} v_i F_i/c$); the second-order terms are similarly relativistic effects arising from the frame transformation. In the space-like component $G_i$, the leading-order term is the radiation force in the comoving frame, while the remaining order $v/c$ and $v^2/c^2$ terms describe relativistic effects that can be interpreted as frame-dragging between matter and radiation.

\subsection{Characteristic numbers and limiting regimes}
\label{ssec:characteristics}

We next derive characteristic dimensionless numbers for RHD systems and consider the limiting behaviour of various terms as we alter their relative sizes. We non-dimensionalise \autoref{eq:hyper} and \autoref{eq:G} following \citet{Lowrie1999}. For the matter quantities, we let $\ell$ be the characteristic size of the system, $a_\infty$ be the characteristic isothermal sound speed, and $\rho_\infty$ be the characteristic density, and define
\begin{gather}
\hx_i = x_i/\ell, \quad \hatt = t / (\ell/a_\infty), \quad \hrho = \rho/\rho_\infty, \quad \hv_i = {v_i}/a_\infty, \\ \hp = p/(\rho_\infty a_\infty^2), \quad \hT = T/T_\infty, \quad \hE_{\rm gas} = E_{\rm gas}/(\rho_\infty a_{\infty}^2),
\end{gather}
Here the quantities with carets are dimensionless versions of the dimensional quantities. 
For the radiation quantities, we let $T_\infty$ be the reference temperature and $\lambda_\infty$ be the reference length scale\footnote{Note that we do not set the reference length scale for radiation quantities to $\ell$ because we wish to ensure that $\hchi_0$ is a quantity of order unity, and $\ell$ and $\lambda_\infty$ may need to be very different in order to accomplish this.}, and we define
\begin{gather}
    \hE = E / (a_r T_{\infty}^4), \quad \hF = F / (c a_r T_{\infty}^4), \quad \hP_{ij} = P_{ij} / (a_r T_{\infty}^4) \\
    \hchi_0 = \chi_0 \lambda_\infty, \quad
4 \pi \hB = 4 \pi B / (c a_r T_\infty^4) = \hT^4,
\end{gather}
With these definitions, the radiation four-force becomes 
\begin{align}
    -cG_0 &= \frac{c a_r T_\infty^4}{\lambda_\infty} \left[
\left(\hchi_{0P} \hT^4 - \hchi_{0E} \hE \right )\left(1 + \frac{1}{2} \frac{\hv^2}{\calC^2}\right) 
\right.
\nonumber \\
& \left.
\qquad
{} + \left(2\hchi_{0E} - \hchi_{0F}\right) \frac{\hv_i \hF_i}{\calC} +
\left(\hchi_{0F} - \hchi_{0E}\right) \frac{\hv^2 \hE + \hv_i \hv_j \hP_{ij}}{\calC^2} \right] \\
&\equiv \frac{c a_r T_\infty^4}{\lambda_\infty} (-\hG_0) \\
    -G_i & = \frac{a_r T_\infty^4}{\lambda_\infty} \left[
-\hchi_{0F} \hF_i \left(1 + \frac{1}{2} \frac{\hv^2}{\calC^2}\right) + \left(\hchi_{0P} \hT^4 - \hchi_{0E} \hE \right) \frac{\hv_i}{\calC}
\right.
\nonumber \\
& \left.
\qquad {} +
\frac{\hchi_{0F}}{\calC} \left(\hv_j \hP_{ji} + \hv_i \hE \right) - \frac{2}{\calC^2}\left(\hchi_{0F}-\hchi_{0E}\right) \hv_j \hF_j \hv_i \right] \\
&\equiv \frac{a_r T_\infty^4}{\lambda_\infty} (-\hG_i),
\end{align}
and the corresponding non-dimensionalised version of \autoref{eq:hyper} is
\begin{equation} \label{eq:rhd_reduced}
\frac{\partial}{\partial\hatt} \hvecU + \hat{\nabla}\cdot \hvecFU = \hvecS_{\hvecU},
\end{equation}
where $\hnabla \equiv \ell \nabla$ and
\begin{equation} \label{eq:F_reduced}
\hvecU =
\left[
\begin{array}{c}
\hrho\\
\hrho\hvecv\\
\hE_{\rm gas}\\
\hE\\
\hvecF
\end{array}
\right], 
\,
\hvecFU =
\left[
\begin{array}{c}
\hrho\hv \\
\hrho\left(\hvecv\otimes \hvecv + \hT\right) \\
\left(\hE_{\rm gas} + \hat{p}\right) \hvecv \\
\calC \hvecF \\
\calC \htenP
\end{array}
\right],
\,
\hvecS_{\hvecU} = \calL
\left[
\begin{array}{c}
0 \\
\calP\hvecG \\
\calP\calC\hG_0 \\
-\calC\hG_0 \\
-\calC\hvecG
\end{array}
\right].
\end{equation}
In these expressions, we have introduced three dimensionless quantities:
\begin{equation}
\calC = \frac{c}{a_\infty}, \qquad
\calL = \frac{\ell}{\lambda_\infty}, \qquad \calP = \frac{a_r T_\infty^4}{\rho_\infty a_\infty^2}.
\end{equation}
These numbers represent, respectively, the dimensionless speed of light, the ratio of the system size to the photon mean free path (and thus is equal to the characteristic optical depth), and (up to factors of order unity) the ratio of radiation pressure to gas pressure at the characteristic temperature and sound speed of the system. It is therefore clear that an RHD system is determined by these three characteristic numbers; the first is always much greater than unity for a non-relativistic system, but the remaining two can be of any size.

Now let us consider various limiting cases of the dimensionless numbers, focusing in particular on the evolution equations for the radiation quantities $\hE$ and $\hvecF$ (the last two entries in \autoref{eq:rhd_reduced}); this will simplify our task since $\calP$ does not appear in these equations. For numerical convenience, and since our goal here is insight rather than rigorous calculation, we will also at this point specialise to the case of grey material, which allows us to choose our scaling $\lambda_\infty$ such that $\hchi_{0P} = \hchi_{0E} = \hchi_{0F} = 1$. In this case the non-dimensional radiation four-force reduces to
\begin{align}
    -\hG_0 &= 
    \left(\hT^4 - \hE\right)\left(1 + \frac{1}{2} \frac{\hv^2}{\calC^2}\right) + \frac{\hv_i \hF_i}{\calC}
    \label{eq:Ghat0} \\
    -\hG_i &= 
    -\hF_i \left(1 + \frac{1}{2} \frac{\hv^2}{\calC^2}\right) + \hT^4 \hv_i + \frac{\hv_j \hP_{ji}}{\calC}. \label{eq:Ghat1}
\end{align}

For $\calL\ll 1$, the system is optically thin and we are in the streaming regime. In this case it is clear that the largest term is $\hat{\nabla}\cdot\hvecF_{\hvecU}$, and so on a fluid flow timescale it is clear that the solution is simply $\hnabla \cdot \hvecF = 0$ and $\hnabla\cdot\htenP = \boldsymbol{0}$. On the other hand, for $\calL \gg 1$, we are in the diffusion limit, and it is clear that $\hvecS_{\hvecU}$, which is of order $\calC\calL$, is the largest term, and therefore on a fluid flow timescale to leading order we must have $(\hG_0, \hvecG) = (0, \boldsymbol{0})$. This in turn requires that, to leading order, $\hE = \hT^4$ and $\hvecF = \boldsymbol{0}$; though we have not shown it here, it is straightforward to show that in this limit we also have $\htenP = (1/3)\tenI \hE$, where $\tenI$ is the identity tensor \citep{Mihalas1984}. Since the flux is zero to the leading order in this case, we must proceed to the next order to determine its value. To do so, we Taylor expand the radiation energy density, flux, and pressure tensor about the leading order solution:
\begin{equation}
    \hE = \hT^4 + \hE_{(1)}
    \quad
    \hvecF = \hvecF_{(1)} 
    \quad
    \htenP = \frac{1}{3}\tenI \hT^4 + \htenP_{(1)},
\end{equation}
where terms subscripted $(1)$ are perturbations that we take to be small compared to the leading-order terms. We insert these expressions into \autoref{eq:rhd_reduced}, Taylor expand, and linearise by dropping all terms that involve products of perturbed quantities. Which terms are at leading order after this procedure depends on the relative values of $\calC$ and $\calL$. If $\calC\gg \calL$, known as the static diffusion limit, then the leading order surviving terms in the equation for the flux are
\begin{equation}
    \frac{\partial}{\partial t}\hvecF_{(1)} + \frac{1}{3} \calC \hnabla \hT^4 = -\calC\calL \hvecF_{(1)}.
\end{equation}
Since both $\calC$ and $\calC\calL$ are large compared to unity, on a fluid flow timescale the terms proportional to these factors must cancel, and therefore we have
\begin{equation}
    \hvecF_{(1)} = -\frac{1}{3\calL}\hnabla\hT^4.
\end{equation}
This is the usual Fick's Law diffusion approximation. Armed with this leading order result, we can see that the relative sizes of the terms in the radiation evolution equations $(\partial/\partial\hatt)(\hE,\hvecF) : \calC\hnabla\cdot (\hvecF,\htenP) : \calC\calL (\hG_0, \hvecG)$ scale relative to one another as $1 : \calC/\calL : \calC /\calL$.

On the other hand, if we have $\calC \ll \calL$, known as the dynamic diffusion case, then the leading non-zero terms are
\begin{equation}
    \frac{\partial}{\partial t}\hvecF_{(1)} = -\calC\calL \hchi_0 \hvecF_{(1)} + \frac{4}{3}\calL \hT^4 \hvecv.
\end{equation}
As in the previous case, since $\calC$ and $\calC\calL$ are large compared to unity, on a fluid flow timescale the two terms on the right hand side must cancel to leading order, and we instead have
\begin{equation}
    \vecF_{(1)} = \frac{4}{3\calC} \hvecv \hT^4.
\end{equation}
In this case the relative scalings of the terms in the radiation evolution equations $(\partial/\partial\hatt)(\hE,\hvecF) : \calC\hnabla\cdot (\hvecF,\htenP) : \calC\calL (\hG_0, \hvecG)$ are $1 : 1 : 1$, i.e., all terms are of equal order.

\subsection{Design considerations for numerical methods}
\label{ssec:rhd_methods}

While the exploration of the limiting cases here is not new \citep[e.g.,][]{Mihalas1984, Lowrie1999, Krumholz2007}, revisiting it allows us to make some important observations about design considerations for numerical methods. First, which terms are large on a fluid flow timescale, and relative to each other, changes from one RHD regime to another -- in the streaming limit the transport terms proportional to $\calC$ dominate, in the static diffusion regime these terms come into balance with the source terms and both are at order $\calC/\calL \gg 1$, while in the dynamic diffusion regime all terms, including the time derivative, become of order unity. Therefore if an RHD scheme is to correctly capture the limiting behaviour in each of these regimes, it cannot rely on any particular assumptions about the relative orderings of these terms, and must be able to cope with situations where each of them is both dominant and sub-dominant.

A second consideration, which we have already introduced in \autoref{sec:intro}, is that in the diffusion regime obtaining the correct solution depends on the dominant terms cancelling to high-order. That is, we have seen that the source terms are naturally of size $\calC\calL$, but in static diffusion the leading order terms cancel and so the dominant non-vanishing term is smaller by a factor $\calL^2 \gg 1$, while for dynamic diffusion it is smaller by a factor $\calC\calL \gg 1$. Similarly, the natural sizes of the transport terms are $\calC$, but this is reduced by a factor $\calL$ in static diffusion, and by a factor $\calC$ in dynamic diffusion. This creates problems for schemes where the transport and source terms, or parts of the source terms, are operator-split, because in an operator split scheme it is difficult to recover the proper near-exact cancellations between the various terms. Overcoming this problem will be our primary objective.

Before describing our proposed solution, however, we pause to discuss another possible approach that is somewhat close to ours in spirit: the DG-IMEX method of \citet{McClarren2008}. This method uses a temporal discretisation that accurately captures cancellations in the static diffusion limit while avoiding non-local implicit solves, and \citeauthor{McClarren2008} show that it passes a number of tests that other operator-split methods fail. However, the price for this is high: the method requires a careful and complex reconstruction scheme that is heavy in terms of both computation and memory usage -- indeed, the method requires eight degrees of freedom per cell, and therefore requires eight times as much memory as the finite volume scheme we present below, a major problem for GPU-based computations where memory is at a premium. Additionally, in the full radiation-hydrodynamics context (not considered by \citeauthor{McClarren2008}), coupling a DG radiation solver to a finite volume hydrodynamics code requires careful consideration of how the fluid internal energy is mapped from the finite volume grid to the DG nodes in order to maintain the asymptotic-preserving property. While existing methods for this mapping numerically manifest the correct asymptotic diffusion solution, they have not been subjected to a rigorous asymptotic analysis such as the one we present below for our scheme, and thus their ability to produce correct asymptotic behaviour over all parameter regimes remains unproven \citep{Bolding2017}. Moreover, extending a DG scheme to adaptive mesh refinement (AMR) would be a substantial challenge, whereas the finite volume scheme we propose is fully compatible with existing AMR frameworks. Finally, \citeauthor{McClarren2008} focus exclusively on the static diffusion regime, and their scheme is not easily extensible to either the streaming or dynamic diffusion limits -- the former because the method relies on a specific closure relation that is a poor approximation for streaming radiation, and the latter because the scheme does not include the velocity-dependent terms that become order unity in the dynamic diffusion limit. Our scheme, by contrast, applies to all RHD regimes.

\section{A New Asymptotic-Preserving Scheme for Radiation Hydrodynamics}
\label{sec:imex}

We now proceed to describe our new numerical method. We first describe our overall strategy for the full RHD system in \autoref{ssec:overall_timestep}, then the IMEX scheme we use for the radiation subsystem in \autoref{ssec:imex_rhd}, and finally our method for carrying out each IMEX stage in \autoref{sec:couple}.

\subsection{Overall time stepping strategy}
\label{ssec:overall_timestep}

We solve the system formed by \autoref{eq:hyper} using an operator split approach consisting of two major steps. In the first step, we advance the hydrodynamic transport subsystem using an explicit method. In the second step, we update the radiation transport subsystem and matter-radiation coupling terms using a mixed implicit-explicit (IMEX) method. 

The hydrodynamic transport subsystem consists of the partial differential equations (PDEs)
\begin{equation}
    \frac{\partial}{\partial t} \left[
    \begin{array}{c}
      \rho \\
      \rho \boldsymbol{v} \\
      E_{\rm gas}
    \end{array}\right] 
    + \nabla \cdot \left[
    \begin{array}{c}
      \rho \boldsymbol{v} \\
      \rho \boldsymbol{v} \otimes \boldsymbol{v}+p \\
      (E_{\rm gas} + p) \boldsymbol{v}
    \end{array}\right]
    = 0.
\end{equation}
Our scheme for radiation does not depend on the numerical method used to solve this subsystem; for the purposes of our implementation in \quokka{} \citep{Wibking2022}, which we use for all the tests below, we adopt a semidiscrete approach, discretising the spatial variables onto a grid while initially keeping the time variable continuous, thereby transforming the partial differential equations into a large set of ODEs. These ODEs are then integrated in time utilizing the second-order accurate, strong stability preserving Runge-Kutta method (RK2-SSP; \citealt{Shu1988}). For an in-depth explanation of the method we direct readers to \cite{Wibking2022}. 

In the present work, our emphasis is on the second major step, the radiation update and matter-radiation coupling. These steps are subcycled with respect to the hydrodynamic step, since they require smaller time steps. 

\subsection{Implicit-explicit method for the radiation subsystem}
\label{ssec:imex_rhd}

Using the same method of lines approach for the radiation subsystem and radiation-matter coupling terms as for hydrodynamics, we define $E_{ijk}$ as the volume-averaged radiation energy density in cell $ijk$, and similarly for all other variables, and express the radiation subsystem for each cell as 
\begin{equation} \label{eq:dysf}
    \frac{d}{dt} \boldsymbol{y}_{ijk} = \boldsymbol{s}(\boldsymbol{y}_{ijk}, t) + \boldsymbol{f}(\boldsymbol{y}_{ijk}, t),
\end{equation}
where, dropping the $ijk$ subscript from this point forward for convenience, we define 
\begin{equation} \label{eq:ysf}
  \vecy = \left[
    \begin{array}{c}
      \rho \boldsymbol{v} \\
      E_{\rm gas} \\
      E \\
      \frac{1}{c^2} \boldsymbol{F}
    \end{array}\right], \quad
    \vecs = - \nabla \cdot \left[
    \begin{array}{c}
      0 \\
      0 \\
      \boldsymbol{F} \\
      \tenP
    \end{array}\right], \quad 
    \vecf =
    \left[
    \begin{array}{c}
      \boldsymbol{G} \\
      c G_0 \\
      - c G_0 \\
      - \boldsymbol{G} 
    \end{array}\right].
\end{equation}
The characteristic time scales associated with the transport term, \(\boldsymbol{s}\), which consists of the divergence of radiation flux, \( \nabla \cdot \boldsymbol{F} \), and the divergence of the radiation pressure tensor \(\nabla \cdot \tenP\), are $\calC^{-1}$, which is potentially fast compared to hydrodynamics, but in the diffusion regime is much longer than the timescale $(\calC\calL)^{-1}$ associated with the source term $\boldsymbol{f}$.\footnote{Of course we have shown above that, due to cancellations, these terms in fact wind up being of the same order -- but our goal is precisely for our numerical method to be able to recover this cancellation.} Consequently, we opt to evolve the transport terms using an explicit method. This choice is strategic; the explicit update obviates the need for global communication across the computational domain. This advantage becomes particularly pronounced on GPUs, where inter-device communication often represents the primary bottleneck in performance. In problems where $\calC$ is so large that this requires infeasibly-many explicit steps, we may elect to solve the approximate RSLA equations instead \citep{Gnedin01a, Skinner13a, Wibking2022}, for which we instead have
\begin{equation}
    \label{eq:yrsla}
  \vecy = \left[
    \begin{array}{c}
      \rho \boldsymbol{v} \\
      E_{\rm gas} \\
      E \\
      \frac{1}{c\hat{c}} \boldsymbol{F}_r
    \end{array}\right], \quad
    \vecs = - \nabla \cdot \left[
    \begin{array}{c}
      0 \\
      0 \\
      \frac{\hat{c}}{c}\boldsymbol{F} \\
      \tenP
    \end{array}\right], \quad 
    \vecf =
    \left[
    \begin{array}{c}
      \boldsymbol{G} \\
      c G_0 \\
      - \hat{c} G_0 \\
      - \boldsymbol{G} 
    \end{array}\right],
\end{equation}
where $\hat{c}$ is the reduced speed of light, chosen so to be $\ll c$ but still much greater than any hydrodynamic speed. This approximation reduces the size of the transport term from $\calC$ to $(\hat{c}/c)\calC$, and thus allows larger time steps. For generality in what follows we will write our update scheme using the RSLA equations, but these can be reduced to the exact equations of RHD simply by setting $\hat{c} = c$.

The short time scales associated with the matter-radiation coupling terms, \( \boldsymbol{f} \), require an implicit treatment. Notably, as these terms do not have spatial derivatives, they allow for the independent update of each cell within the computational domain. Therefore, in the whole radiation update, there is no non-local implicit update, effectively eliminating the requirement for additional inter-domain communication beyond what is standard in a pure hydrodynamic update. 

To ensure that our choice to operator-split between the source and transport terms in this manner does not compromise accuracy, we utilize the asymptotic-preserving IMEX PD-ARS integrator \citep{Chu2019}, a choice motivated by its proven effectiveness in preserving the diffusion limit while maintaining second-order accuracy and stability in the streaming limit.\footnote{Formally, \citet{Chu2019} show that IMEX PD-ARS reduces to RK2-SSP when $\boldsymbol{f}$ is small compared to $\boldsymbol{s}$, which is the case in the streaming limit.} The IMEX PD-ARS integrator can be characterized by its double Butcher tableau
\begin{multline}
{\cal B}_{\rm PD-ARS} = \\
\begin{array}{c|ccc}
0 & 0 & 0 & 0 \\
1 & 1 & 0 & 0 \\
1 & 1/2 & 1/2 & 0 \\
\hline
& 1/2 & 1/2 & 0
\end{array}
~~~~~~
\begin{array}{c|ccc}
0 & 0 & 0 & 0 \\
1 & 0 & 1 & 0 \\
1 & 0 & 1/2 - \epsilon & 1/2 + \epsilon \\
\hline
& 0 & 1/2 - \epsilon & 1/2 + \epsilon,
\end{array}
\end{multline}
where $\epsilon$ is a free parameter in the range $[0,0.5)$; in our implementation we adopt $\epsilon = 0$.
When expressed in equations, the scheme to advance the system by a time $\Delta t$ is
\begin{align}
    \vecy^{(n+1/2)} =& ~\vecy^{(n)} + \Delta t ~\vecs (\vecy^{(n)}) + \Delta t ~\vecf (\vecy^{(n+1/2)}, t + \Delta t) \label{eq:imex1} \\
    \vecy^{(n+1)} =& ~\vecy^{(n)} + \Delta t \left[ \frac{1}{2} ~\vecs(y^{(n)}) + \frac{1}{2} ~\vecs(y^{(n+1/2)}, t + \Delta t) \right] \nonumber \\
    &+ \Delta t \left[ \frac{1}{2} ~\vecf(y^{(n+1/2)}, t + \Delta t) + \frac{1}{2} ~\vecf(y^{(n+1)}, t + \Delta t) \right], \label{eq:imex2}
\end{align}
where the superscript $(n)$ indicates the state at the start of the radiation update (but after the operator-split hydrodynamic update), $(n+1)$ indicates the state at the end of the radiation update, and $(n+1/2)$ indicates an intermediate stage.

One key feature of this update scheme is that the transport and source terms appear symmetrically at each of the two stages, so that cancellations can be captured properly. A second key feature is that, on modern architectures where communication is expensive, this scheme is only marginally more costly than an update like RK2-SSP, because the only excess work it requires is an extra iterative solve at the first stage (\autoref{eq:imex1}). Crucially, this extra solve is purely local, because only the local source terms, rather than the non-local transport terms, must be iterated. Thus there is no extra communication in this scheme relative to RK2-SSP.

\subsection{Update strategy for IMEX stages} 
\label{sec:couple}

At each stage of the integrator above, the right-hand side contains both explicit terms -- those that are evaluated at state $(n)$ during the first IMEX stage (\autoref{eq:imex1}) and at states $(n)$ or $(n+1/2)$ during the second stage (\autoref{eq:imex2}) and implicit terms that are evaluated at state $(n+1/2)$ during the first stage and state $(n+1)$ during the second stage. We therefore begin each stage by evaluating the explicit terms. This in turn requires that we evaluate the transport terms $\vecf$, which are always explicit in keeping with our overall strategy. We do this using a Godunov method. Specifically, we compute the fluxes for $E$ and $\vecF$ between cells by using PPM reconstruction to obtain the states at the cell edges and then using a Harten, Lax, van Leer (HLL) Riemann solver \citep{Harten1983} to compute the fluxes from these states. The details of this step are elaborated in \cite{Wibking2022}, and hence we will not repeat them here. However, we modify the scheme in an important way compared to the method presented in \citeauthor{Wibking2022}: that scheme, and all previous comparable explicit treatments of radiation, have been forced to invoke an \textit{ad hoc} correction to the wavespeeds computed in the Riemann solver in an attempt to recover the diffusion limit. With our new time-stepping scheme, no such modification is required, and we can simply use the uncorrected HLL fluxes in our update. We discuss this issue further in \autoref{ssec:wavespeeds}.

After carrying out the explicit part of the update, we are left with the implicit part. For both stages of the IMEX integrator we can express this stage in the generic form 
\begin{gather}\label{eq:egupdate}
    E_{\rm gas}^{(t+1)} - E_{\rm gas}^{(t)} = c G_0^{(t+1)} \theta \Delta t \\
    \label{eq:eupdate}
    E^{(t+1)} - E^{(t)} = - \chat G_0^{(t+1)} \theta \Delta t
\end{gather}
for the energies, and 
\begin{gather}
(\rho \vecv)^{(t+1)} - (\rho \vecv)^{(t)} = \vecG^{(t+1)} \theta\Delta t \label{eq:vupdate} \\
\vecF^{(t+1)} - \vecF^{(t)} = - \chat c \vecG^{(t+1)} \theta\Delta t \label{eq:fupdate}
\end{gather}
for the momenta. Here terms with superscript \((t)\) denote the state after applying the explicit terms -- for example during the first stage (\autoref{eq:imex1}) we have $E^{(t)} = E^{(n)} - \Delta t (\hat{c}/c) \nabla\cdot \vecF^{(n)}$ -- and terms with superscript \((t+1)\) denote the final state for which we are attempting to solve; the factor $\theta$ is unity during the first stage and $1/2$ during the second stage.

In each cell this is a system of eight equations in eight unknowns -- radiation and gas energy, three components of gas velocity, and three components of radiation flux -- that must be solved simultaneously. While we could do so straightforwardly using a single Newton-Raphson iteration scheme or similar, it is more efficient to separate the iteration procedure into an inner stage where we solve the energy equations while freezing $\vecv$ and $\vecF$, and a second, outer stage where update the flux and gas velocity and then if necessary go back to the inner stage and recompute $E$ and $E_{\rm gas}$ using the updated values of $\vecv$ and $\vecF$. The reason this is more efficient is that it allows the inner iteration stage to consider only two variables rather than eight, and in most cases the work terms proportional to $\vecv$ and $\vecF$ are small in the energy equations, so $E$ and $E_{\rm gas}$ change little to none as a result of the update to $\vecv$ and $\vecF$ and the whole procedure converges in a single or at most a few outer iterations.

The inner stage consists of updating the energy quantities using a modified version of the iteration scheme initially proposed by \citet{Howell2003} and modified by \citet{Wibking2022}. We solve \autoref{eq:egupdate} and \autoref{eq:eupdate} for $E_{\rm gas}^{(t+1)}$ and $E^{(t+1)}$ by performing Newton-Raphson iteration on the non-linear implicit system
\begin{gather} \label{eq:FGFR}
    0 = F_G(E_{\rm gas}^{(t+1)}, E^{(t+1)}) \equiv E_{\rm gas}^{(t+1)} - E_{\rm gas}^{(t)} + \left( \frac{c}{\chat} \right) R^{(t+1)}, \\
    0 = F_R(E_{\rm gas}^{(t+1)}, E^{(t+1)}) \equiv E^{(t+1)} - E^{(t)} - (R^{(t+1)} + S^{(t+1)}),
\end{gather}
where 
\begin{equation} \label{eq:Rnew}
    \begin{split}
    R^{(t+1)} &\equiv - \chat G_0^{(t+1)} \theta\Delta t \\
    &= \theta\Delta t \left[\chat \left(\chi_{0P} \frac{4 \pi B}{c} - \chi_{0E} E\right)\left(1 + \frac{1}{2} \frac{v^2}{c^2}\right)\right. \\
    & \qquad {} + (2\chi_{0E} - \chi_{0F}) \frac{\chat}{c} \left(\frac{v_i F_i}{c} \right)
    \\ 
    & \qquad \left. {}+ \hat{c} (\chi_{0F} - \chi_{0E}) \left(\betasq E + \frac{v_i v_j P_{ij}}{c^2}\right)\right],
    \end{split}
\end{equation}
and $S$ is an optional term to include, for example, the addition of radiation by stellar sources. Note that $E$, $\tenP$, and all the variables that can depend on $E_{\rm gas}$ -- $\chi_{0P}$, $\chi_{0E}$, $\chi_{0F}$, and $B$ -- carry the superscripts ${(t+1)}$ which are omitted here for the sake of brevity. By contrast, we freeze $\vecv$ and $\vecF$ at their values at the start of the inner stage as noted above.

A single Newton-Raphson iteration consists of solving the linearised equations 
\begin{equation}
  \label{eq:linear}
  \mathbf{J} \cdot \Delta \vecx = - \vecF(\vecx),
\end{equation}
where $\vecx$ is the set of variables to be updated, $\Delta\vecx$ is the change in these variables during this iteration, $\vecF(\vecx)$ is the vector whose zero we wish to find, and $\mathbf{J}$ is the Jacobian matrix of \( \vecF(\vecx) \). Instead of taking $\vecx = (E_{\rm gas}, E)$, as in the original \citet{Howell2003} scheme, we use $\vecx = (E_{\rm gas}, R)$ as the base variables over which to iterate; we find that the system generally converges in fewer iterations using this basis, likely because at high optical depth the system almost immediately converges to the solution $R = 0$, and thus the remaining iterations are effectively on $E_{\rm gas}$ alone The Jacobian in this basis is 
\begin{equation}
\begin{split}
    \frac{\partial F_G}{\partial E_{\rm gas}} &= 1 \\
    \frac{\partial F_G}{\partial R} &= \frac{c}{\hat{c}} \\
    \frac{\partial F_{R}}{\partial E_{\rm gas}} &= \left.\frac{\partial E}{\partial E_{\rm gas}}\right|_{R = {\rm const.}} = \frac{1}{C_v} \frac{\partial}{\partial T} \left( \frac{\chi_{0P}}{\chi_{0E}} \frac{4 \pi B}{c} \right) \\
    \frac{\partial F_{R}}{\partial R} &= \left.\frac{\partial E}{\partial R}\right|_{T = {\rm const.}} - 1 = - \frac{1}{\hat{c} \ \chi_{0E} \ \theta\Delta t } - 1.
\end{split}
\end{equation}
We have omitted the $v^2/c^2$ terms and assumed $\partial (\kappa_P/\kappa_E)/\partial T = 0$ in the calculation of the Jacobian for simplicity, but this simplification only changes the rate of convergence; it does not affect the converged solution. 

After solving \autoref{eq:linear} for $\Delta \vecx$, we update $\vecx \leftarrow \vecx + \Delta \vecx$; we do this repeatedly until the system converges, as determined by the condition
\begin{equation}
    \left| \frac{F_G}{E_{\rm tot}} \right| < \epsilon \quad {\rm and} \quad \left| \frac{c}{\chat} \frac{F_R}{E_{\rm tot}} \right| < \epsilon
\end{equation}
where
\begin{equation}
    E_{\rm tot} = E_{\rm gas}^{(t)} + \frac{\chat}{c} (E_r^{(t)} + S^{(t)}).
\end{equation}
is the total radiation and material energy at the beginning of the timestep accounted for reduced speed of light. We set the relative tolerance $\epsilon = 10^{-13}$ by default. Once this Newton-Raphson system converges, we have the updated gas total energy \(E_{\rm gas}^{(t+1)}\) and we can compute the updated radiation energy as 
\begin{equation}
E^{(t+1)} = E^{(t)} - \frac{c}{\hat{c}}(E_{\rm gas}^{(t+1)} - E_{\rm gas}^{(t)}) + S^{(t+1)}.
\end{equation}
Note that, for $\hat{c} = c$ and $S = 0$, this procedure ensures that our scheme conserves total energy to machine precision regardless of the level of accuracy with which we have iterated the equations to convergence.

We then proceed to the outer stage of the iteration where we solve the flux and momentum update equations, \autoref{eq:vupdate} and \autoref{eq:fupdate}, with the updated gas temperature, opacity, and radiation energy. To order $v/c$, the solution is straightforward:
\begin{eqnarray}\label{eq:numF1}
    \lefteqn{F_i^{(t+1)} = } 
    \\
    & & \frac{F_i^{(t)} +  \hat{c} \theta \Delta t\left[ \chi_{0P} \frac{4 \pi B}{c} v_i + \chi_{0F} v_j P_{ji} + (\chi_{0F} - \chi_{0E}) E v_i \right]}{1 + \chat \chi_{0F} \theta\,\Delta t}.
    \nonumber
\end{eqnarray}
To order $v^2/c^2$, in cases where $\chi_{0F} - \chi_{0E} = 0$, the solution is
\begin{equation}\label{eq:numF2}
    F_i^{(t+1)} = \frac{F_i^{(t)} + \hat{c} \theta\Delta t  \left( \chi_{0P} \frac{4 \pi B}{c} v_i + \chi_{0F} v_j P_{ji} \right)}{1 + \chat \chi_{0F} (1 + v^2/2c^2) \theta\Delta t }.
\end{equation}
When $\chi_{0F} \ne \chi_{0E}$ and to order $v^2/c^2$, the solution is slightly more complex because all three components of $F$ appear in $G_i$. In this case, $F_i^{(n+1)}$  is the solution of a set of three linear equations; these are straightforward to solve analytically, but the resulting expressions are somewhat lengthy and so we omit them here for brevity. Lastly, following \autoref{eq:vupdate}, we update the gas momentum via 
\begin{equation}
    (\rho v_i)^{(t+1)} = (\rho v_i)^{(t)} - \frac{F_i^{(t+1)} - F_i^{(t)}}{c \chat}.
\end{equation}
This update also ensures momentum conservation to machine precision. After we update the gas momentum, we also recalculate the gas's internal energy (which in \textsc{Quokka} we track separately because we implement a dual energy formalism) by subtracting the updated kinetic energy from the updated total energy.

As previously indicated, the gas velocity $\vecv$ and radiation flux $\vecF$ we use in the inner stage of the iteration are lagged. This can cause significant inaccuracies at high optical depths when the velocity-dependent terms are non-negligible. To eliminate errors like this and render this scheme fully implicit, we now repeat the inner iteration using the updated values of $\vecv$ and $\vecF$, and compute new estimates for $\vecv$ and $\vecF$, repeating this procedure until either the relative change in the value of the terms proportional to $\vecv$ and $\vecF$ in $R$ (\autoref{eq:Rnew}) from one outer iteration to the next  is below $10^{-13}$ or the absolute change is below $10^{-13} R$. 
Except in the dynamic diffusion limit, where the velocity-dependent terms are at the same order as all other terms, this iterative process typically terminates after just one iteration, and in all the tests we present below, and for all the test problems presented in \cite{Wibking2022}, we never require more than a handful of outer iterations. Thus the cost is modest. This completes accounting for all terms in the radiation four-force, thus completing radiation-matter coupling.

\section{Properties of the scheme in the diffusion limit} 
\label{sec:asymp}

Before proceeding to numerical tests of the scheme we have described, we first present an analysis of its behaviour in the so-called asymptotic diffusion regime, where the photon mean free path is not resolved by the computational grid, in order to demonstrate directly why it succeeds in capturing this limit while other schemes fail.

\subsection{The asymptotic diffusion limit of the discrete IMEX equations} 

We begin our analysis by recalling the results from \autoref{ssec:characteristics}, which are that in the static diffusion limit for grey material with $\hchi_{0E} = \hchi_{0F} = \hchi_{0P} = 1$, to leading order the radiation energy in non-dimensionalised variables is
\begin{equation}\label{eq:e0}
\hE = \hT^4
\end{equation}
and the radiation flux is
\begin{equation} \label{eq:f1}
    \hvecF = - \frac{1}{3 \calL} \hnabla \hT^4.
\end{equation}
Inserting these limits into the evolution equations for matter and radiation energy that we solve during the radiation update (\autoref{eq:rhd_reduced}; i.e., omitting changes in the matter energy due to fluid processes), we have
\begin{align}
    \frac{\partial\hE_\mathrm{gas}}{\partial\hatt} &= \calP \calC \hG_0 \\
    \frac{\partial\hE}{\partial\hatt} &= \frac{\calC}{3\calL} \hnabla^2 \hT^4 - \calC \hG_0,
\end{align}
and thus the evolution equation for the total matter plus radiation energy reduces to the usual radiation-diffusion form,\footnote{Note that the total energy in dimensionless units is $\hE + \calP^{-1} \hE_\mathrm{gas}$ rather than $\hE + \hE_\mathrm{gas}$ because in our non-dimensionalisation the matter and radiation energies are scaled differently -- the matter energy is normalised by $\rho_\infty a_\infty^2$, while the radiation energy is normalised by $a_r T_\infty^4$, and these two differ by a factor of $\calP$. See \autoref{ssec:characteristics} for details.}
\begin{equation}
    \frac{\partial}{\partial \hatt}\left(\hE + \calP^{-1}\hE_\mathrm{gas}\right) = \frac{\calC}{3\calL} \hnabla^2 \hT^4.
    \label{eq:diffu}
\end{equation}
For a numerical method for thermal radiative transfer to be ``asymptotic preserving'', it must give a valid discretization of \autoref{eq:diffu} and enforce the conditions in \autoref{eq:e0} and \autoref{eq:f1} when $\calC \gg \calL \gg 1$. With an asymptotic preserving method, it is possible to use cells that are optically thick and still obtain accurate solutions of radiative transfer.

To verify that our IMEX scheme satisfies this condition, we begin by writing down the two steps of the IMEX update for the radiation energy (\autoref{eq:imex1} and \autoref{eq:imex2}), again using non-dimensionalised variables and assuming grey material. For simplicity we will adopt $\hvecv = 0$ and $\hat{c} = c$ as well. This gives
\begin{multline}
    \frac{\hE^{(n+1/2)} - \hE^{(n)}}{\Delta \hatt} = \\
    - \hnabla\cdot (\calC \hvecF^{(n)}) - {\cal L} \calC \left[\hE^{(n+1/2)} - (\hT^{(n+1/2)})^4 \right], \label{eq:Ea}
\end{multline}
followed by
\begin{multline}
     \frac{\hE^{(n+1)} - \hE^{(n)}}{\Delta \hatt} = - \frac{1}{2} \left[ \hnabla\cdot (\calC \hvecF^{(n)}) + \hnabla\cdot (\calC \hvecF^{(n+1/2)}) \right] \\
     - \frac{1}{2} {\cal L} \calC \left[ \hE^{(n+1/2)} - (\hT^{(n+1/2)})^4 + \hE^{(n+1)} - (\hT^{(n)})^4 \right], \label{eq:Eb}
\end{multline}
From \autoref{eq:Ea} we can solve for $E^{(n+1/2)}$ and expand to first order in ${\cal L}^{-1}$ to obtain
\begin{equation}
    \hE^{(n+1/2)} = (\hT^{(n+1/2)})^4 + {\cal L}^{-1} \left[ (\calC \Delta \hatt)^{-1} \hE^{(n)} - \hnabla\cdot \hvecF^{(n)} \right].
\end{equation}
Thus in limit $\calC\gg \calL \gg 1$, we have $\hE^{(n+1/2)} = (\hT^{(n+1/2)})^4$. Similarly, performing the same operation on \autoref{eq:Eb}, one can show to leading order in $\calL^{-1}$ that
\begin{equation}\label{eq:pr_e0}
    \hE^{(n+1)} = (T^{(n+1)})^4.
\end{equation}
We have therefore established the our scheme enforces \autoref{eq:e0} at both stages of the IMEX update.

We next examine the two IMEX stages for the radiation flux,\footnote{Note that in writing down these equations we assume that our closure for the pressure tensor will produce $\tenP \to (1/3)\tenI E$ in optically thick conditions, which is true of any reasonable closure scheme.}
\begin{align}
    \frac{\hvecF^{(n+1/2)} - \hvecF^{(n)}}{\Delta \hatt} = & \ - \hnabla \left(\frac{\calC E^{(n)}}{3}\right) - {\cal L} \calC \hvecF^{(n+1/2)} \label{eq:Fa} \\
    \frac{\hvecF^{(n+1)} - \hvecF^{(n)}}{\Delta \hatt} = & \ - \frac{1}{2} \left[ \hnabla \left(\frac{\calC \hE^{(n)}}{3}\right) + \hnabla \left(\frac{\calC \hE^{(n+1/2)}}{3}\right) \right] \nonumber \\
    & \ - \frac{1}{2} {\cal L} \calC (\hvecF^{(n+1/2)} + \hvecF^{(n+1)}). \label{eq:Fb}
\end{align}
\autoref{eq:Fa} implies 
\begin{equation} \label{eq:ff1}
    \vecF^{(n+1/2)} = - \frac{1}{1 + {\cal L} \calC \Delta \hatt} \left( \frac{\calC \Delta \hatt}{3} \hnabla\hE^{(n)} - \vecF^{(n)} \right)
\end{equation}
and if we again take the limit $\calC\gg\calL\gg 1$, to leading order we have
\begin{equation}
     \hvecF^{(n+1/2)} = - \frac{1}{3\calL} \hnabla\hE^{(n)} = - \frac{1}{3\calL} \hnabla(\hT^{(n)})^4.
     \label{eq:flux_nhalf}
\end{equation}
Applying the same procedure to \autoref{eq:Fb} yields leading-order terms
\begin{equation}
    \label{eq:ff2}
    \hvecF^{(n+1)} = - \frac{1}{3\calL}\hnabla \hE^{(n+1/2)} = - \frac{1}{3\calL}\hnabla (\hT^{(n+1/2)})^4.
\end{equation}
Thus the leading term of the radiation flux reduces the form given by \autoref{eq:f1}, simply lagged by a half-step. This nonetheless means that our scheme obeys this constraint.

Finally, we write down the two IMEX stages for the matter energy update
\begin{align}\label{eq:Ga}
    \calP^{-1} \frac{\hE_{\rm gas}^{(n+1/2)} - \hE_{\rm gas}^{(n)}}{\Delta \hatt} &= {\cal L} \calC \left[ \hE^{(n+1/2)} - (\hT^{(n+1/2)})^4 \right], \\
    \calP^{-1} \frac{\hE_{\rm gas}^{(n+1)} - \hE_{\rm gas}^{(n)}}{\Delta \hatt} &= 
    \frac{1}{2} {\cal L} \calC \left[ \hE^{(n+1/2)} - (\hT^{(n+1/2)})^4 \right.
    \nonumber \\
    & \left.\qquad {} + E^{(n+1)} - (T^{(n)})^4 \right],
    \label{eq:Gb}
\end{align}
Adding \autoref{eq:Ea} to \autoref{eq:Ga} and using \autoref{eq:flux_nhalf} we get 
\begin{align}
    & \frac{\hE^{(n+1/2)} - \hE^{(n)}}{\Delta \hatt} + \calP^{-1} \frac{\hE_{\rm gas}^{(n+1/2)} - \hE_{\rm gas}^{(n)}}{\Delta \hatt} 
    \nonumber \\
    & \quad\quad = - \hnabla \cdot (\calC \hvecF^{(n)}) 
    \nonumber \\
    & \quad\quad = \frac{\calC}{3\calL} \hnabla^2 (\hT^{(n-1/2)})^4,
    \label{eq:EEa}
\end{align}
where the superscript $(n-1/2)$ indicates the intermediate state of the \textit{previous} time step. Similarly, adding \autoref{eq:Eb} to \autoref{eq:Gb} and using \autoref{eq:ff2} gives 
\begin{align}
    &\frac{\hE^{(n+1)} - \hE^{(n)}}{\Delta \hatt} + \calP^{-1} \frac{\hE_{\rm gas}^{(n+1)} - \hE_{\rm gas}^{(n)}}{\Delta \hatt} \nonumber \\
    &\quad \quad = - \frac{1}{2} \left( \hnabla\cdot(\calC \hvecF^{(n)}) + \hnabla\cdot(\calC \hvecF^{(n+1/2)}) \right) \nonumber \\
    &\quad \quad= \frac{\calC}{3\calL} \left(\frac{1}{2}\right) \left[ \hnabla^2(\hT^{(n)})^4 + \hnabla^2 (\hT^{(n+1/2)})^4 \right]. \label{eq:EEb}
\end{align}
The combination of \autoref{eq:EEa} and \autoref{eq:EEb} represents a valid discretisation of the diffusion equation \autoref{eq:diffu}, albeit one using a temperature that is lagged by half a step, thus proving our IMEX PD-ARS scheme preserves the asymptotic diffusion limit. 

It is worth noting that schemes such as the one described by \autoref{eq:EEa} and \autoref{eq:EEb}, while they represent valid discretisations of the diffusion equation, can be unstable depending on how the spatial derivatives are evaluated. In particular, \citet{Radice2018} point out that, in a scheme where computation of $\hvecF$ in the diffusion limit effectively reduces to computing a centred difference on $\hE$, and in turn evaluating $\hnabla\cdot\hvecF$ in the equations above reduces to evaluating a centred difference on $\hvecF$, the resulting scheme has the property that the solutions in even- and odd-numbered cells are decoupled, i.e., the solution in cell $i$ depends only on the states in cells $i-2$ and $i+2$, not $i-1$ and $i+1$; this in turn can give rise to an even-odd instability where numerical oscillations with a period of two cells are not damped and can grow large. \citeauthor{Radice2018} propose a method to suppress this instability. In our tests, while we observe faint hints of the instability in our scheme, these appear only in the dynamic diffusion regime and only in tests at very low resolution (e.g., $<64$ cells per linear dimension). In all other problems the instability, if it exists at all, is imperceptibly small. We therefore do not use \citeauthor{Radice2018}'s correction for any of the tests we present below. However, we have implemented it in \quokka{}, and allow users to enable it via a compile-time option should it prove useful at some point in the future.

\subsection{Comparison to schemes with purely explicit intermediate stages}

It is instructive at this point to repeat the analysis we have just performed for the IMEX discretisation for the RK2-SSP discretisation used in \citet{Wibking2022}, since this will let us see why this scheme does not successfully capture the asymptotic diffusion regime. While our analysis will be specific to this particular time stepping approach, we will see that the results generalise straightforwardly to any scheme where the intermediate stage is fully explicit and includes only the transport terms, and thus to other schemes such as those proposed by \citet{Rosdahl2015} and \citet{Skinner2019}.

The RK2-SSP scheme uses the time update
\begin{align}
    \boldsymbol{y}^{(n+1/2)} = \boldsymbol{y}^{(n)} &+ \Delta t ~\boldsymbol{s}(\boldsymbol{y}^{(n)}, t) \\
    \boldsymbol{y}^{(n+1)} = \boldsymbol{y}^{(n)} &+ \Delta t \left[ \frac{1}{2} ~\boldsymbol{s}(\boldsymbol{y}^{(n)}, t) + \frac{1}{2} ~\boldsymbol{s}(\boldsymbol{y}^{(n+1/2)}, t+\Delta t) \right] \nonumber 
    \\
    & {}
    + \Delta t \boldsymbol{f}(\boldsymbol{y}^{(n+1)}, t + \Delta t)
\end{align}
Compared to \autoref{eq:imex1} and \autoref{eq:imex2}, the primary difference is in the treatment of the fast terms $\boldsymbol{f}$; in the IMEX scheme these appear at both stages of the update, while in the RK2-SSP scheme they appear only at the final stage. This will prove to be crucial in what follows.

Again adopting the limit $\calC \gg \calL \gg 1$ and setting $\tenP = (1/3) \tenI E$, as appropriate for static diffusion, we can write down the leading-order terms in the two stages of this update for the radiation energy as
\begin{gather}
    \hE^{(n+1/2)} = \hE^{(n)} - \calC \Delta \hatt \hnabla\cdot \hvecF^{(n)} \\
    \hE^{(n+1)} = (\hT^{(n+1)})^4.
\end{gather}
For the radiation flux, we obtain
\begin{gather}
    \hvecF^{(n+1/2)} = \hvecF^{(n)} - \frac{\calC \Delta \hatt}{3} \hnabla \hE^{(n)}, \label{eq:ssp-f1} \\
    \hvecF^{(n+1)} = -\frac{1}{3\calL} \left(\frac{1}{2}\right)\left( \hnabla \hE^{(n)} + \hnabla \hE^{(n+1/2)}\right). \label{eq:ssp-f2}
\end{gather}
Finally, if we write down the update for the gas energy (which is non-trivial only for the second stage, since $s = 0$ for the gas energy) and add it to that for the radiation energy, 
\begin{multline}
    \frac{\hE^{(n+1)} - \hE^{(n)}}{\Delta \hatt} + \calP^{-1}\frac{\hE_{\rm gas}^{(n+1)} - \hE_{\rm gas}^{(n)}}{\Delta \hatt} = \\
    - \frac{\calC}{2} \left( \hnabla\cdot \hvecF^{(n)} + \hnabla\cdot \hvecF^{(n+1/2)} \right). \label{eq:ea2}
\end{multline}
Substituting \autoref{eq:ssp-f1} and \autoref{eq:ssp-f2} into \autoref{eq:ea2}, we obtain
\begin{multline}
\frac{\hE^{(n+1)} - \hE^{(n)}}{\Delta \hatt} + \calP^{-1}\frac{\hE_{\rm gas}^{(n+1)} - \hE_{\rm gas}^{(n)}}{\Delta \hatt} = \\
\frac{1}{2} \left[ \left(\frac{1}{2}\right) \frac{\calC}{3\calL} (\hnabla^2\hE^{(n)} + \hnabla^2 \hE^{(n-1/2)}) + \frac{\calC^2 \Delta \hatt}{3} \hnabla^2 \hE^{(n)} \right].
\label{eq:EErk2}
\end{multline}
Thus we see that there are two modes of radiation diffusion in this numerical scheme: one is physical diffusion with a diffusion coefficient of $\calC/3\calL$, and the other is numerical diffusion with a coefficient of $\calC^2 \Delta \hatt/3$. This numerical mode will dominate for any time step $\Delta \hatt \gtrsim 1/\calC\calL$, or, in dimensional terms, $\Delta t > \lambda_\infty/c$, i.e., whenever the time step is large enough that we do not resolve the light-crossing time of a photon mean free path. In the asymptotic diffusion regime, where the photon mean free path is smaller than the size of a cell, this means that numerical diffusion dominates any time that the time step is larger than the light crossing time of a cell. In practice, the time step is always much larger than this -- since otherwise one might as well use a fully explicit scheme -- which explains why discretisations such as RK2-SSP fail in the asymptotic diffusion regime.

Comparing \autoref{eq:EErk2} to \autoref{eq:EEb}, we see that the numerical diffusion mode is removed in the IMEX discretisation, and by comparing the calculations leading up to these equations we can also understand why. The numerical diffusion term in the RK2-SSP update originates in a term that appears at the intermediate stage of the flux update (\autoref{eq:ssp-f1}). This term does not appear to leading order in the intermediate stage of the IMEX update (\autoref{eq:flux_nhalf}) because it is overwhelmed by the source term, which is a factor of $\calL$ larger; it is a cancellation within the source term that forces the radiation flux to the correct value for diffusion. Thus the IMEX update winds up with an estimate for the intermediate-state flux that is of order $1/\calL$ independent of the time step, while the RK2-SSP update, because it ignores the order $\calC\calL$ source term but retains the order $\calC$ transport term during the intermediate stage, obtains a flux estimate that is of order $\calC \Delta \hatt$ instead. This overestimate is what gives rise to the artificial numerical diffusion of energy. An important conclusion to draw from this argument is that the failing in the RK2-SSP scheme for this problem is not specific to that scheme, but is instead generic to \textit{any} update scheme that contains a stage that includes only the transport terms and not the source terms. Such a scheme will always overestimate the transport in diffusion-regime problems where the source term is responsible for suppressing them.

\subsection{On modifications to the radiation wave speed}
\label{ssec:wavespeeds}

The problem we have identified in the RK2-SSP scheme is not new; indeed, multiple authors have pointed out that the HLL solver applied to the radiation subsystem, in a scheme where the source and transport terms are operator-split, yields fluxes that fail to preserve the asymptotic diffusion limit \citep{Lowrie2001,Audit2002,Jiang2013,Skinner2019,Wibking2022}. In an attempt to circumvent this problem, these authors have proposed a range of corrections to the energy fluxes; for instance, \citet{Skinner2019} suggest
\begin{equation}
\mathcal{F}_{E, \text { corrected }}^{\mathrm{HLL}}=\frac{S_R F_L-S_L F_R+\epsilon S_R S_L\left(E_R-E_L\right)}{S_R-S_L}
\end{equation}
where $S_{L,R}$ are the characteristic left and right wavespeeds, respectively, whose values absent the RSLA are given by $S = \pm\sqrt{f} c$, where $f$ is the component of the Eddington tensor along that direction. In this expression, $\epsilon$ is an empirical correction factor that smoothly transitions from $1$ in the streaming limit to $1/\tau$ in the optically thick limit, where $\tau$ is an empirical estimate of the optical depth, usually computed as the optical depth across a handful of computational cells. This correction implies an effective wave speed of 
\begin{equation}\label{eq:Sjiang}
S_{\rm eff} = c \sqrt{\frac{f}{\tau}},
\end{equation}
In light of the preceding discussion, we can see that this reduction in the flux is effectively a correction that attempts to reduce the numerical diffusion mode in \autoref{eq:EErk2}, by in turn forcing the overestimated intermediate time-flux (\autoref{eq:ssp-f1}) back toward the solution that would have been obtained by retaining rather than dropping the source term. However, the accuracy of this fix, particularly in the regime of intermediate optical depth, is not known.

In our IMEX PD-ARS method, we use the wavespeed $S$ without any correction in the Riemann solver. No correction is necessary because, by retaining the source term, we automatically recover the correct flux in the diffusion limit (\autoref{eq:flux_nhalf}), and our scheme correctly and smoothly goes between the optically thin limit, where the transport term dominates, and the optically thick limit, where the source term does. Indeed, a corollary of this analysis is that, in the highly optically thick regime where the source term is dominant, we need not even obey the CFL condition for the radiation in order to obtain the correct answer. To understand why this is the case, we notice that \autoref{eq:EEb} is an effective temporal discretisation of the diffusion equation
\begin{equation}\label{eq:diff1}
    \frac{\partial}{\partial \hatt} \hE = \frac{\calC}{3 \calL} \hnabla^2 \hT^4,
\end{equation}
which has a diffusion coefficient of $\hat{D}_{\rm diff} = \calC / (3\calL)$. The effective speed of the diffusion in time interval $\hatt$ is approximately
\begin{equation}
    \hat{S}_{\rm diff} \approx \frac{\sqrt{\hat{D}_{\rm diff} \hatt}}{\hatt} = \sqrt{\frac{\calC}{3 \calL \hatt}}
\end{equation}
To better understand the magnitude of $\hat{S}_{\rm diff}$, we reintroduce dimensions to the variables and get
\begin{equation}
    S_{\rm diff} = \sqrt{\frac{c}{3 t}} = \frac{c}{\sqrt{3}} \frac{1}{\sqrt{{\rm CFL} ~\tau_{\rm cell}}}
\end{equation}
where CFL $\equiv c t / \Delta x$ is the radiation CFL number. This evaluation of diffusion speed aligns with the wave speed correction `hack' (\autoref{eq:Sjiang}) when CFL = 1. Our IMEX scheme's asymptotic preserving nature, as demonstrated, allows the CFL number to exceed 1 while still accurately capturing the diffusion limit. This is exemplified in the subsequent discussion on the asymptotic Marshak wave test (\autoref{sec:marshak}), where our numeric scheme accurately catches the position of the diffusion front for radiation CFL numbers up to $\sim 10$. This observation is in concordance with \cite{Jiang2013}, who, through a linear analysis of the diffusion equations, showed that if one calculates the effective propagation speed with a wavelength of 10 cells, the numerical diffusion is small enough not to affect the solution. This compatibility with the findings from linear wave analysis of diffusion equations not only validates our scheme's robustness in limiting numerical diffusion effectively but also underscores the versatility of the IMEX PD-ARS scheme in accommodating higher CFL numbers without compromising the asymptotic preserving property.

\section{Numerical Tests}
\label{sec:test}

The original \quokka{} paper \citep{Wibking2022} introduced a series of tests to validate the accuracy and convergence capabilities of \quokka{}, including tests of the hydrodynamic subsystem, the radiation transport, and coupled RHD. This new scheme passes all these tests, so we will not repeat them here. We instead introduce additional tests for our implementation of the IMEX PD-ARS scheme in \quokka{} that are specifically designed to test the code's ability to preserve the asymptotic diffusion limit. The full source code and outputs for all tests are available in the \quokka{} github repository (see Data Availability statement for details).

\subsection{Non-equilibrium Radiation Shock} 

\begin{figure}
    \centering
    \includegraphics[width=\columnwidth]{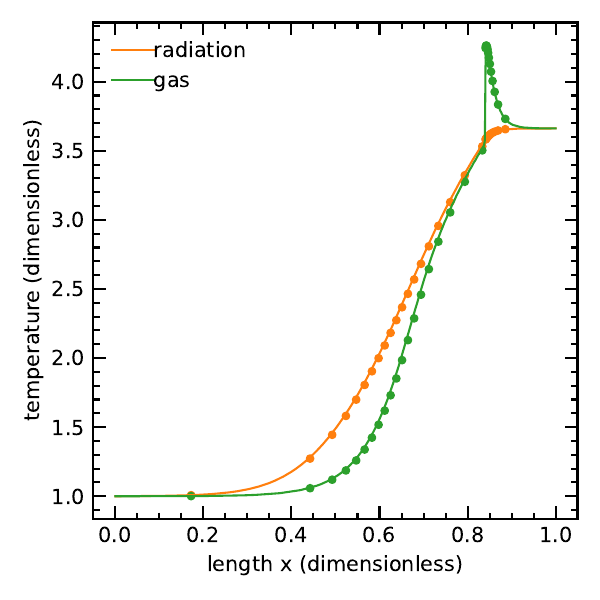}    
    \caption{Comparing Radiation and gas temperature from the numerical calculation (solid lines) with the exact steady-state solution (dots) in a subscritical radiation shock with $M = 3$.}
    \label{fig:shock}
\end{figure}

We begin our tests with the classical grey non-equilibrium radiative shock test described by \cite{Zeldovich1967}. 
Radiation can modify the structure of a shock because it diffuses and interacts with matter. \cite{Lowrie2008} have found a semi-analytic exact solution of radiative shocks with grey -non-equilibrium diffusion. 
Using their parameters with an upstream Mach number of ${\cal M} = 3$, we obtain a subcritical radiation shock whose temperature jumps discontinuously at the shock interface. We present our numerical calculation of this problem and compare them to the solutions of \cite{Lowrie2008}. Following the setup used by \cite{Skinner2019}, we scale from dimensionless to cgs units by setting the opacities to $\kappa = 577 \ {\rm cm^2~g^{-1}}$ and mean molecular weight to $\mu = m_{\rm H}$. We use an adiabatic EOS with an adiabatic index to $\gamma = 5/3$. The shock is simulated in a 1D region with $x \in [0, 0.01575]$ cm resolved with 512 grids. The problem is set in the rest frame of the shock initialised at $x=0$, and the conditions on the left and right sides of the shock are uniform, with densities, temperatures, and velocities given by $\rho_L = 5.69 {\rm g~cm^{-3}}, T_L = 2.18 \times 10^6 K, v_L = 5.19 \times 10^7 {\rm cm~s^{-1}}$, and  $\rho_R = 17.1 \ {\rm g~cm^{-3}}, T_R = 7.98 \times 10^6 \ K, v_R = 1.73 \times 10^7 \ {\rm cm~s^{-1}}$, respectively. In order to exactly match the assumptions used in the semi-analytic solution, we use the Eddington approximation, $\tenP = (1/3) E \tenI$, to calculate the radiation pressure tensor. Following \cite{Skinner2019}, we use a reduced speed of light $\hat{c} = 10 (v_L + c_{s,L})$, where $c_{s, L}$ is the adiabatic sound speed of the left-side state. We use a CFL number of 0.4 and evolve until $t=10^{-9}$ s. We show the resulting temperature profile in \autoref{fig:shock}. The agreement between our numerical calculation and the semi-analytic solution is excellent. The $L_1$ norm of the relative error of the gas temperature is 0.38 per cent, which is as good as the solution of \cite{Skinner2019}. 

\subsection{The Asymptotic Marshak wave problem}\label{sec:marshak}

We test the code's ability to accurately capture radiation in the optically thick limit via the diffusive Marshak-wave problem. This problem consists of a semi-infinite medium of material with a variable absorption coefficient $\chi = 300\mbox{ cm}^{-1} / (k_B T/\mbox{keV})^3$. We input incoming isotropic radiation on the boundary at $x=0$ from a 1 keV temperature source. We use an initial condition of $T(x) = T_0 = 10^{-3}$ keV and $E(x) = a_r T_0^4$ for our test problem. 

\begin{figure*}
    \centering
    \includegraphics[width=0.49\textwidth]{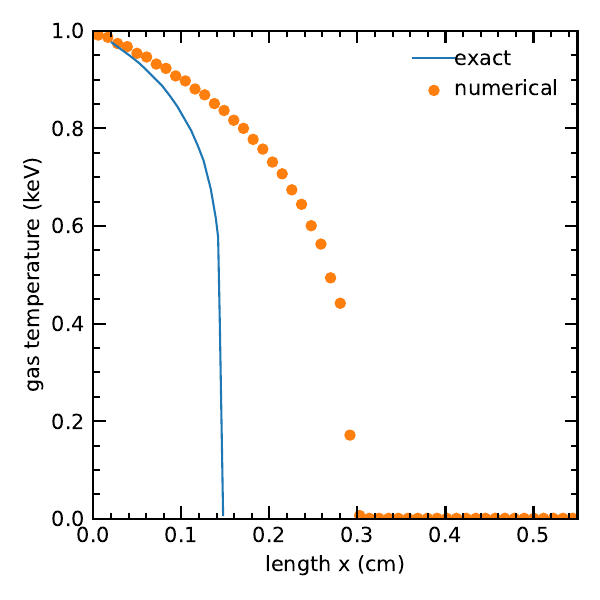}\includegraphics[width=0.5\textwidth]{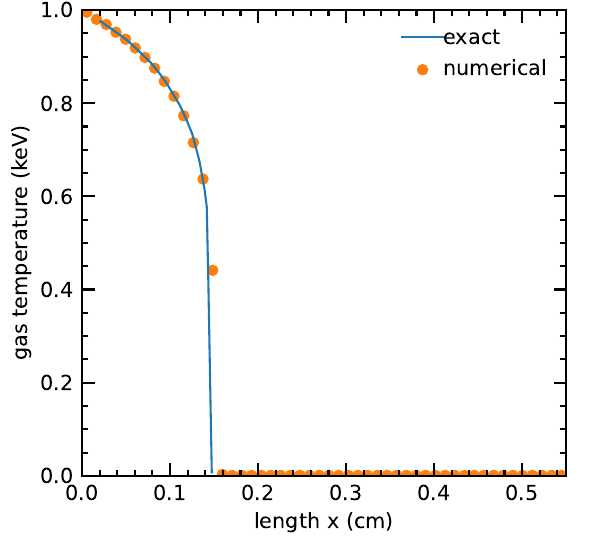}
    \caption{Comparing numerical solutions to the Marshak-wave problem in the asymptotic diffusion limit from the SSP-RK2 scheme (left) and IMEX PD-ARS scheme we introduce here (right). The solid lines are the semi-analytic solutions and the dots are the nodal values of the numerical solution. Both simulations run with $\Delta x = 0.011$ cm and CFL $=0.9$. The IMEX scheme accurately captures the diffusion of radiation while in the SSP-RK2 scheme solution, the wavefront propagates about two times too fast.}
    \label{fig:marshak}
\end{figure*}

\autoref{fig:marshak} shows our numerical results compared to the semi-analytic solution for the diffusive Marshak-wave problem, alongside a comparison to the solution using the SSP-RK2 scheme. The numerical calculation uses a spatial grid of 60 cells across the domain (0, 0.66 cm), with each cell spanning $3 \times 10^9$ mean-free paths at the minimum temperature and 3 mean-free paths at the highest temperature. The heat capacity in this problem is constant at $3 \times 10^{15} \ {\rm erg~cm^{-3}~keV^{-1}}$. We compare this solution with the semi-analytic equilibrium-diffusion solution from \cite{Zeldovich2002}. We find a relative L1-norm error at about $4.5 \%$. The simulation results presented here are obtained using a CFL number of 0.9, but the simulation runs with a CFL number up to $\sim 10$, implying a time step roughly 10 times the largest permissible time step in the streaming limit. 

\subsection{Advecting radiation pulse in the static diffusion limit} 

\begin{figure*}
    \centering 
    \includegraphics[width=0.49\textwidth]{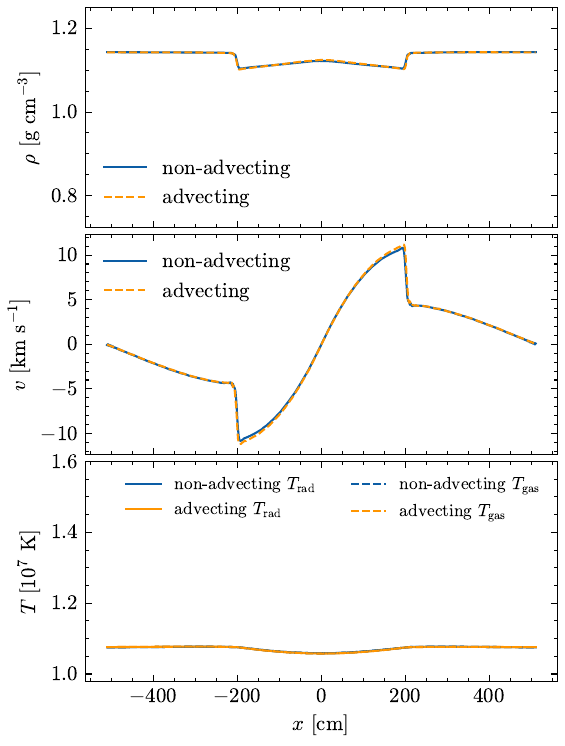}
    \includegraphics[width=0.48\textwidth]{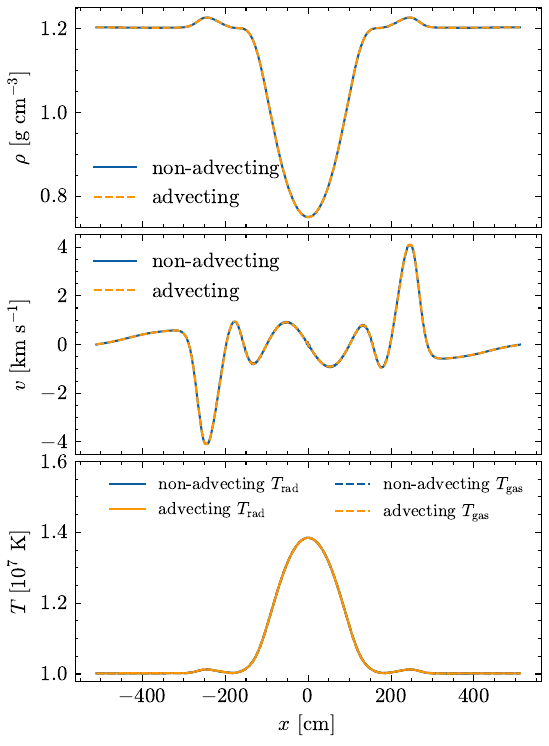}
    \caption{Advecting radiation pulse tests in the static diffusion regime at $t = 4.8 \times 10^{-5}$ s using the RK2-SSP scheme (left) versus our newly introduced IMEX scheme (right). The IMEX scheme accurately captures the radiation diffusion, agreeing well with the results presented in \protect\cite{Krumholz2007} and \protect\cite{Zhang11a}.}
    \label{fig:pulse}
\end{figure*}

\begin{figure}
    \centering
    \includegraphics[width=\columnwidth]{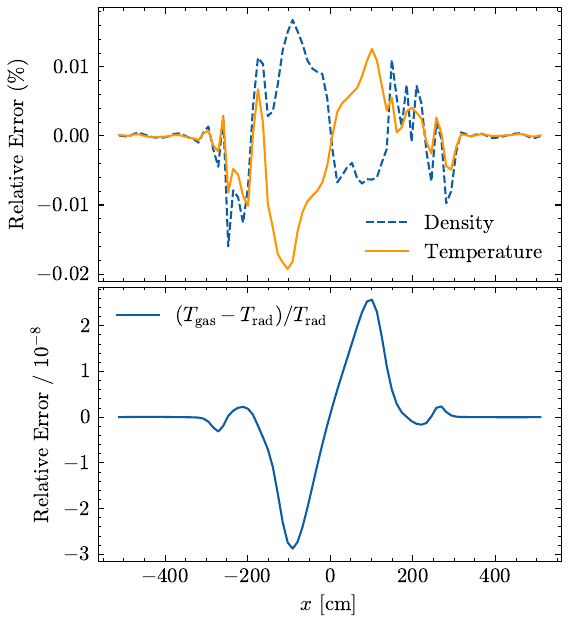}
    \caption{Top: Relative errors in density (dashed lines) and temperature (solid lines) between the advected and unadvected runs. Bottom: Relative difference between gas and radiation temperature in the advected run.}
    \label{fig:pulse-error}
\end{figure}

In this test, introduced by \cite{Krumholz2007}, we simulate the advection of a radiation pulse by the gas motion in the diffusion regime. We set the initial gas and radiation temperatures equal in this test, and the initial temperature and density profiles are 
$$
\begin{gathered}
T=T_0+\left(T_1-T_0\right) \exp \left(-\frac{x^2}{2 w^2}\right), \\
\rho=\rho_0 \frac{T_0}{T}+\frac{a_R \mu}{3 k_{\mathrm{B}}}\left(\frac{T_0^4}{T}-T^3\right),
\end{gathered}
$$
with $T_0=10^7 \mathrm{~K}, T_1=2 \times 10^7 \mathrm{~K}, \rho_0=1.2 \mathrm{~g} \mathrm{~cm}^{-3}, w=24 \mathrm{~cm}$, and $\mu=2.33 m_p=3.9 \times 10^{-24} \mathrm{~g}$. The radiation pressure is estimated via the Eddington approximation. The opacity of the gas is set at $\kappa_{0P} = \kappa_{0E} = \kappa_{0F} = 100 \ {\rm cm^2~g^{-1}}$. The system is initially in pressure balance. If there were no radiation diffusion, the system would be in an equilibrium between the gas pressure and radiation pressure. Because of radiation diffusion, the balance is lost and the gas moves. 

We solve the problem numerically in two different frames, one in the lab frame and the other in the comoving frame, and compare the results in \autoref{fig:pulse}. In the comoving frame run, the velocity is 0 in the beginning everywhere. In the lab frame run, the initial velocity is $v_0 = 10^6 \ {\rm cm~s^{-1}}$. In both runs, the optical depth across the pulse is $\tau = \rho \kappa w = 2.9 \times 10^3$, and in the lab-frame run $\beta \equiv v_0/c = 3.3 \times 10^{-5}$, giving $\beta \tau \approx 0.1$ and placing this problem in the static diffusion limit. 
The computational domain in both runs is a 1D region of  ($-512$ cm, 512 cm) with periodic boundaries, and the grid consists of 512 uniform cells. 
We show the density, temperature, and velocity profiles from both runs at $t=4.8 \times 10^{-5}$ s in \autoref{fig:pulse}. The results of the lab-frame run have been shifted in space by $v_0 t$ for comparison. The density, temperature, and velocity profiles from both runs are almost identical, demonstrating the accuracy of our scheme in capturing radiation advection. We also show the relative difference from both runs in \autoref{fig:pulse-error}. The relative difference between the non-advecting and advecting cases is below $0.03 \%$ anywhere. The relative difference between the temperature of the gas and radiation is below $3 \times 10^{-9}$. We compare these results with that of \cite{Krumholz2007} and \cite{Zhang11a} and find good agreement. 

\subsection{Advecting radiation pulse in the dynamic diffusion limit}

\begin{figure}
    \centering
    \includegraphics[width=\columnwidth]{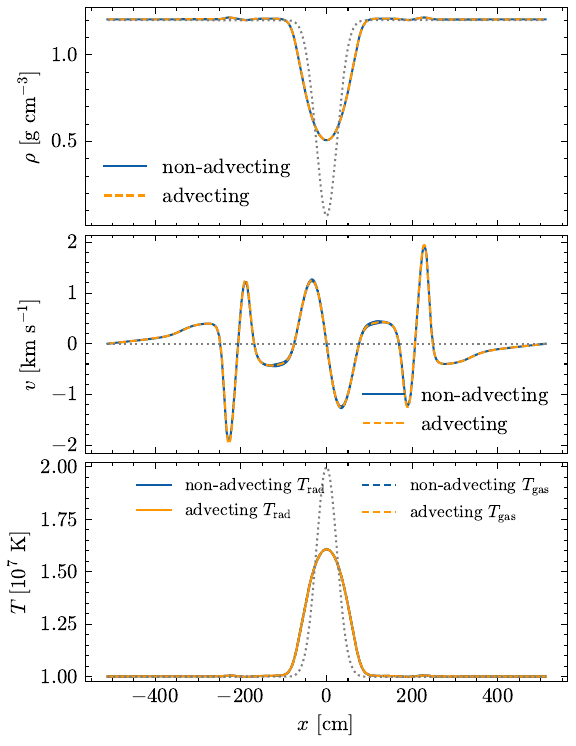}
    \caption{Advecting radiation pulse test in the dynamic diffusion regime, where $\beta \equiv v/c = 10^{-3}$ and the optical depth across the pulse is $\tau = 10^4$, yielding $\beta \tau = 10$. The initial condition is plotted as grey dotted lines. The observed radiation diffusion closely matches the theoretical predictions derived from the diffusion equation. The difference between the advected and non-advected pulse is so small that it is nearly invisible to the eye.}
    \label{fig:pulse-dyn}
\end{figure}

In this test, we rerun the advecting radiation pulse problem in the dynamic diffusion regime. Compared to the static diffusion case, the advection speed is increased to $3 \times 10^7 \ {\rm cm~s^{-1}}$ and the opacity is increased to $500 \ {\rm cm^2~g^{-1}}$. Numerically, $\beta = 10^{-3}$, $\tau = 1.4 \times 10^4$, and $\beta \tau = 14$, placing this problem in the dynamic diffusion limit. We increase the number of cells to 1024 to reduce the magnitude of odd-even decoupling instability, and evolve the simulation to $t = 4.8 \times 10^{-5}$ s. The relative difference between the non-advecting and advecting cases is below 0.06 per cent anywhere (\autoref{fig:pulse-dyn}), demonstrating the excellent accuracy of our scheme in capturing radiation advection in the dynamic diffusion regime. Although there is no analytic solution to this problem, we can estimate the expected pulse width via the diffusion equation 
\begin{equation}
    \frac{\partial}{\partial t} E = \frac{c}{3\chi} \frac{\partial^2}{\partial x^2} E,
\end{equation}
The expected distance that the pulse has diffused at time $t$ is approximately
\begin{equation}
    w \approx \sqrt{\frac{c t}{3\chi}} = 28.3 ~{\rm cm},
\end{equation}
The temperature profile of our numerical calculation \autoref{fig:pulse-dyn} agrees well with this analytic estimation.

\subsection{Advecting uniform medium in the dynamic diffusion regime}

\begin{figure}
    \centering
    \includegraphics[width=\columnwidth]{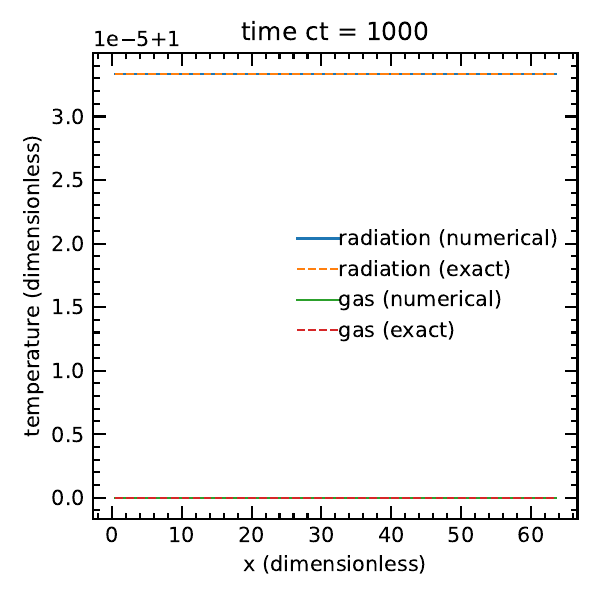}
    \caption{Advecting uniform medium test in the dynamic diffusion limit ($\beta = 0.01$, $\beta \tau = 10^3$). The numerical results match the analytic solution to the limits of machine precision.}
    \label{fig:uniform}
\end{figure}

\begin{table}
    \centering 
    \begin{tabular}{llll}
$\beta$&  $\beta \tau$  &Order&Error \\
\hline
$10^{-4}$&
1&1&$< 10^{-15}$\\
$10^{-2}$& $10^3$& 1&$4 \times 10^{-5}$\\
$10^{-2}$& $10^3$& 2&$< 10^{-15}$\\
    \end{tabular}
    \caption{Relative errors of the matter temperature in numerical calculation with respect to theoretical solution in the advecting uniform medium tests. Col. 3: the $v/c$ order used in the source terms.}
    \label{tab:adv}
\end{table}

We test the code's ability to incorporate relativistic corrections up to order $v/c$ or order $v^2/c^2$ with a 1-D test problem of advecting a uniform medium in the dynamic diffusion regime. In this test, the gas is initially moving at a uniform velocity of $v_0 = 0.01 c$ and has a uniform density. We configure the opacity and grid size such that each grid's optical depth is $10^5$. The radiation is initially in thermal equilibrium with the medium at the comoving frame. 

In the lab frame, to order $v/c$ the initial radiation quantities are 
\begin{align}
    & E(t=0) = E_0, \\
    & F(t=0) = \frac{4}{3} v_0 E_0,
\end{align}
and we can show that the space-like component of the radiation four-force, \autoref{eq:G1}, to order $v/c$ vanishes: 
\begin{align}
  -G_1 & =-\chi_{0} \frac{{F}}{c} + \chi_{0} \frac{4 \pi B}{c} \frac{v}{c} + \chi_{0} \frac{{v_j} P_{j1}}{c} \\
  & = -\chi_{0} \frac{4}{3} E_0 \frac{v_0}{c} + \chi_{0} E_0 \frac{v_0}{c} + \chi_{0} \frac{1}{3} E_0 \frac{v}{c} \\
  & = 0
\end{align}

To order $v^2/c^2$ the initial radiation quantities are
\begin{align}
    & E(t=0) = E_0 \left(1 + \frac{4}{3} \frac{v^2}{c^2}\right), \\
    & F(t=0) = \frac{4}{3} v_0 E_0,
\end{align}
and we can show that both the time-like and space-like component of the radiation four-force vanishes:
\begin{align}
    -c G_0 & =c \chi_0 \left(\frac{4 \pi B}{c} - E\right)\left(1+\frac{1}{2}\betasq\right) + \chi_{0} \frac{v_i F_i}{c}
  \nonumber \\
  & = c \chi_0 \left[ E_0 - E_0 (1 + \frac{4}{3} \frac{v^2}{c^2}) \right] \left(1+\frac{1}{2}\betasq\right) + \chi_{0} \frac{v_0}{c} \frac{4}{3} v_0 E_0 \nonumber \\
  & = O\left(\frac{v^4}{c^4}\right)
\end{align}
\begin{align}
  -G_1 & =-\chi_{0F} \frac{{F}}{c} \left(1+\frac{1}{2} \betasq \right) + \chi_{0P} \frac{4 \pi B}{c} \frac{v}{c} + \chi_{0F} \frac{{v_j} P_{j1}}{c}
  \nonumber \\ 
  & = -\chi_{0F} \frac{4}{3} E_0 \frac{{v_0}}{c} \left(1+\frac{1}{2} \frac{v_0^2}{c^2} \right) + \chi_{0P} E_0 \frac{v_0}{c} \nonumber \\
  {} & \qquad +\chi_{0F} \frac{{v_0}}{c} \frac{1}{3} E_0 (1 + \frac{4}{3} \frac{v_0^2}{c^2}) \nonumber \\
  & = O\left(\frac{v^3}{c^3}\right)
\end{align}

The key feature of this test is that, in the lab frame, the radiation energy is out of equilibrium with the thermal radiation. However, this excess is offset by the work done to the radiation by the matter. One can show that our backwards Euler scheme, \autoref{eq:numF1} and \autoref{eq:numF2}, preserves this equilibrium, leading to an equilibrium state ($E^{(n+1)} = E^{(n)}$ and $\vecF^{(n+1)} = \vecF^{(n)}$) that aligns precisely with the expected outcomes of this test (\autoref{fig:uniform}), achieving precision up to machine accuracy even in the dynamic diffusion limit (\autoref{tab:adv}). Conversely, the source term from \cite{Howell2003} and \cite{Wibking2022}, defined as $\propto (4 \pi B/c - E_r)$, fails to attain this balance.

\section{Conclusion} 

We have presented a novel mixed implicit-explicit (IMEX) time integration scheme for finite-volume RHD that is second-order accurate in the streaming limit and accurately preserves the asymptotic diffusion limit. Our method uniquely combines the robustness of {\it local} implicit methods in handling stiff source terms with the scalability and simplicity of explicit methods for advective transport, making it particularly suitable for adaptive mesh refinement and massively parallel, GPU-accelerated architectures. 
This scheme addresses a critical gap in current RHD solvers by ensuring asymptotic accuracy in the streaming, static diffusion, and dynamic diffusion limits, including in the asymptotic diffusion regime where the photon mean free path is much smaller than the computational grid, without necessitating non-local implicit steps or \textit{ad hoc} adjustments to the radiation flux based on the optical depth. We have implemented our algorithm in the GPU-accelerated AMR RHD code \quokka{} \citep{Wibking2022}. \quokka{} is open-source and a link to download the source code is in the DATA AVAILABILITY section.

We have verified our algorithm using a variety of established quantitative tests, including the non-equilibrium radiation shock test, the asymptotic Marshak wave test, and the advecting radiation pulse test in the static and dynamic diffusion limit. These tests demonstrate the scheme's capability to recover the correct asymptotic limits. 

Looking forward, we envision several areas for further development. Extending our scheme to include more complex radiation transport models, such as multi-frequency radiation transfer and non-thermal emission, could enhance its applicability to a broader range of astrophysical problems. Additionally, integrating our RHD method with other physics modules, such as magnetic fields, could open new channels for multi-physics simulations in astrophysics.

\section*{Acknowledgements}

CCH and MRK acknowledge support from the Australian Research Council through Laureate Fellowship FL220100020. This research was undertaken with the assistance of resources and services from the National Computational Infrastructure (NCI) and the Pawsey Supercomputing Centre, which are supported by the Australian Government.

\section*{Data Availability}
 
The \quokka{} code, including the source code and summary result plots for all tests presented in this paper, is available from \url{https://github.com/quokka-astro/quokka} under an open-source license.

\bibliographystyle{mnras}
\bibliography{refs}

\begin{thebibliography}{}
\makeatletter
\relax
\def\mn@urlcharsother{\let\do\@makeother \do\$\do\&\do\#\do\^\do\_\do\%\do\~}
\def\mn@doi{\begingroup\mn@urlcharsother \@ifnextchar [ {\mn@doi@} {\mn@doi@[]}}
\def\mn@doi@[#1]#2{\def\@tempa{#1}\ifx\@tempa\@empty \href {http://dx.doi.org/#2} {doi:#2}\else \href {http://dx.doi.org/#2} {#1}\fi \endgroup}
\def\mn@eprint#1#2{\mn@eprint@#1:#2::\@nil}
\def\mn@eprint@arXiv#1{\href {http://arxiv.org/abs/#1} {{\tt arXiv:#1}}}
\def\mn@eprint@dblp#1{\href {http://dblp.uni-trier.de/rec/bibtex/#1.xml} {dblp:#1}}
\def\mn@eprint@#1:#2:#3:#4\@nil{\def\@tempa {#1}\def\@tempb {#2}\def\@tempc {#3}\ifx \@tempc \@empty \let \@tempc \@tempb \let \@tempb \@tempa \fi \ifx \@tempb \@empty \def\@tempb {arXiv}\fi \@ifundefined {mn@eprint@\@tempb}{\@tempb:\@tempc}{\expandafter \expandafter \csname mn@eprint@\@tempb\endcsname \expandafter{\@tempc}}}

\bibitem[\protect\citeauthoryear{Audit, Charrier, Chi{\`e}ze  \& Dubroca}{Audit et~al.}{2002}]{Audit2002}
Audit E.,  Charrier P.,  Chi{\`e}ze J.~P.,   Dubroca B.,  2002, \mn@doi [arXiv e-prints] {10.48550/arXiv.astro-ph/0206281}, pp astro--ph/0206281

\bibitem[\protect\citeauthoryear{{Bolding}, {Hansel}, {Edwards}, {Morel}  \& {Lowrie}}{{Bolding} et~al.}{2017}]{Bolding2017}
{Bolding} S.,  {Hansel} J.,  {Edwards} J.~D.,  {Morel} J.~E.,   {Lowrie} R.~B.,  2017, \mn@doi [Journal of Computational Physics] {10.1016/j.jcp.2017.02.063}, \href {https://ui.adsabs.harvard.edu/abs/2017JCoPh.338..511B} {338, 511}

\bibitem[\protect\citeauthoryear{Chu, Endeve, Hauck  \& Mezzacappa}{Chu et~al.}{2019}]{Chu2019}
Chu R.,  Endeve E.,  Hauck C.~D.,   Mezzacappa A.,  2019, \mn@doi [Journal of Computational Physics] {10.1016/j.jcp.2019.03.037}, 389, 62

\bibitem[\protect\citeauthoryear{{Davis} \& {Tchekhovskoy}}{{Davis} \& {Tchekhovskoy}}{2020}]{Davis2020}
{Davis} S.~W.,  {Tchekhovskoy} A.,  2020, \mn@doi [\araa] {10.1146/annurev-astro-081817-051905}, \href {https://ui.adsabs.harvard.edu/abs/2020ARA&A..58..407D} {58, 407}

\bibitem[\protect\citeauthoryear{{Gnedin} \& {Abel}}{{Gnedin} \& {Abel}}{2001}]{Gnedin01a}
{Gnedin} N.~Y.,  {Abel} T.,  2001, \mn@doi [\na] {10.1016/S1384-1076(01)00068-9}, \href {http://adsabs.harvard.edu/abs/2001NewA....6..437G} {6, 437}

\bibitem[\protect\citeauthoryear{Harten, Lax  \& van Leer}{Harten et~al.}{1983}]{Harten1983}
Harten A.,  Lax P.~D.,   van Leer B.,  1983, \mn@doi [SIAM Review] {10.1137/1025002}, 25, 35

\bibitem[\protect\citeauthoryear{He, Ricotti  \& Geen}{He et~al.}{2019}]{He2019}
He C.-C.,  Ricotti M.,   Geen S.,  2019, \mn@doi [Monthly Notices of the Royal Astronomical Society] {10.1093/mnras/stz2239}, 489, 1880

\bibitem[\protect\citeauthoryear{Howell \& Greenough}{Howell \& Greenough}{2003}]{Howell2003}
Howell L.~H.,  Greenough J.~A.,  2003, \mn@doi [Journal of Computational Physics] {10.1016/S0021-9991(02)00015-3}, 184, 53

\bibitem[\protect\citeauthoryear{{Jiang}, {Stone}  \& {Davis}}{{Jiang} et~al.}{2012}]{Jiang12a}
{Jiang} Y.-F.,  {Stone} J.~M.,   {Davis} S.~W.,  2012, \mn@doi [\apjs] {10.1088/0067-0049/199/1/14}, \href {http://adsabs.harvard.edu/abs/2012ApJS..199...14J} {199, 14}

\bibitem[\protect\citeauthoryear{Jiang, Stone  \& Davis}{Jiang et~al.}{2013}]{Jiang2013}
Jiang Y.-F.,  Stone J.~M.,   Davis S.~W.,  2013, \mn@doi [The Astrophysical Journal] {10.1088/0004-637X/767/2/148}, 767, 148

\bibitem[\protect\citeauthoryear{Krumholz, Klein, McKee  \& Bolstad}{Krumholz et~al.}{2007}]{Krumholz2007}
Krumholz M.~R.,  Klein R.~I.,  McKee C.~F.,   Bolstad J.,  2007, \mn@doi [ApJ] {10.1086/520791}, 667, 626

\bibitem[\protect\citeauthoryear{{Krumholz}, {Klein}, {McKee}, {Offner}  \& {Cunningham}}{{Krumholz} et~al.}{2009}]{Krumholz09c}
{Krumholz} M.~R.,  {Klein} R.~I.,  {McKee} C.~F.,  {Offner} S.~S.~R.,   {Cunningham} A.~J.,  2009, \mn@doi [Science] {10.1126/science.1165857}, \href {http://adsabs.harvard.edu/abs/2009Sci...323..754K} {323, 754}

\bibitem[\protect\citeauthoryear{Lowrie \& Edwards}{Lowrie \& Edwards}{2008}]{Lowrie2008}
Lowrie R.~B.,  Edwards J.~D.,  2008, \mn@doi [Shock Waves] {10.1007/s00193-008-0143-0}, 18, 129

\bibitem[\protect\citeauthoryear{{Lowrie} \& {Morel}}{{Lowrie} \& {Morel}}{2001}]{Lowrie2001}
{Lowrie} R.~B.,  {Morel} J.~E.,  2001, \mn@doi [\jqsrt] {10.1016/S0022-4073(00)00097-2}, \href {https://ui.adsabs.harvard.edu/abs/2001JQSRT..69..475L} {69, 475}

\bibitem[\protect\citeauthoryear{Lowrie, Morel  \& Hittinger}{Lowrie et~al.}{1999}]{Lowrie1999}
Lowrie R.~B.,  Morel J.~E.,   Hittinger J.~A.,  1999, \mn@doi [The Astrophysical Journal] {10.1086/307515}, 521, 432

\bibitem[\protect\citeauthoryear{McClarren, Evans, Lowrie  \& Densmore}{McClarren et~al.}{2008}]{McClarren2008}
McClarren R.~G.,  Evans T.~M.,  Lowrie R.~B.,   Densmore J.~D.,  2008, \mn@doi [Journal of Computational Physics] {10.1016/j.jcp.2008.04.029}, 227, 7561

\bibitem[\protect\citeauthoryear{Menon, Federrath, Krumholz, Kuiper, Wibking  \& Jung}{Menon et~al.}{2022}]{Menon2022}
Menon S.~H.,  Federrath C.,  Krumholz M.~R.,  Kuiper R.,  Wibking B.~D.,   Jung M.,  2022, \mn@doi [Monthly Notices of the Royal Astronomical Society] {10.1093/mnras/stac485}, 512, 401

\bibitem[\protect\citeauthoryear{{Menon}, {Federrath}  \& {Krumholz}}{{Menon} et~al.}{2023}]{Menon22c}
{Menon} S.~H.,  {Federrath} C.,   {Krumholz} M.~R.,  2023, \mn@doi [\mnras] {10.1093/mnras/stad856}, \href {https://ui.adsabs.harvard.edu/abs/2023MNRAS.521.5160M} {521, 5160}

\bibitem[\protect\citeauthoryear{Mihalas}{Mihalas}{1978}]{Mihalas1978}
Mihalas D.,  1978, Stellar Atmospheres.
Astronomy and Astrophysics Series, W. H. Freeman

\bibitem[\protect\citeauthoryear{Mihalas \& Mihalas}{Mihalas \& Mihalas}{1984}]{Mihalas1984}
Mihalas D.,  Mihalas B.~W.,  1984, Foundations of Radiation Hydrodynamics.
Oxford University Press, \url {https://ui.adsabs.harvard.edu/abs/1984oup..book.....M}

\bibitem[\protect\citeauthoryear{{Naab} \& {Ostriker}}{{Naab} \& {Ostriker}}{2017}]{Naab2017}
{Naab} T.,  {Ostriker} J.~P.,  2017, \mn@doi [\araa] {10.1146/annurev-astro-081913-040019}, \href {https://ui.adsabs.harvard.edu/abs/2017ARA&A..55...59N} {55, 59}

\bibitem[\protect\citeauthoryear{{Radice}, {Abdikamalov}, {Ott}, {M{\"o}sta}, {Couch}  \& {Roberts}}{{Radice} et~al.}{2018}]{Radice2018}
{Radice} D.,  {Abdikamalov} E.,  {Ott} C.~D.,  {M{\"o}sta} P.,  {Couch} S.~M.,   {Roberts} L.~F.,  2018, \mn@doi [Journal of Physics G Nuclear Physics] {10.1088/1361-6471/aab872}, \href {https://ui.adsabs.harvard.edu/abs/2018JPhG...45e3003R} {45, 053003}

\bibitem[\protect\citeauthoryear{Rosdahl \& Teyssier}{Rosdahl \& Teyssier}{2015}]{Rosdahl2015}
Rosdahl J.,  Teyssier R.,  2015, \mn@doi [Monthly Notices of the Royal Astronomical Society] {10.1093/mnras/stv567}, 449, 4380

\bibitem[\protect\citeauthoryear{{Rosen}, {Krumholz}, {McKee}  \& {Klein}}{{Rosen} et~al.}{2016}]{Rosen16a}
{Rosen} A.~L.,  {Krumholz} M.~R.,  {McKee} C.~F.,   {Klein} R.~I.,  2016, \mn@doi [\mnras] {10.1093/mnras/stw2153}, \href {http://adsabs.harvard.edu/abs/2016MNRAS.463.2553R} {463, 2553}

\bibitem[\protect\citeauthoryear{Shu \& Osher}{Shu \& Osher}{1988}]{Shu1988}
Shu C.-W.,  Osher S.,  1988, \mn@doi [Journal of Computational Physics] {10.1016/0021-9991(88)90177-5}, 77, 439

\bibitem[\protect\citeauthoryear{{Skinner} \& {Ostriker}}{{Skinner} \& {Ostriker}}{2013}]{Skinner13a}
{Skinner} M.~A.,  {Ostriker} E.~C.,  2013, \mn@doi [\apjs] {10.1088/0067-0049/206/2/21}, \href {http://adsabs.harvard.edu/abs/2013ApJS..206...21S} {206, 21}

\bibitem[\protect\citeauthoryear{{Skinner}, {Burrows}  \& {Dolence}}{{Skinner} et~al.}{2016}]{Skinner2016}
{Skinner} M.~A.,  {Burrows} A.,   {Dolence} J.~C.,  2016, \mn@doi [\apj] {10.3847/0004-637X/831/1/81}, \href {https://ui.adsabs.harvard.edu/abs/2016ApJ...831...81S} {831, 81}

\bibitem[\protect\citeauthoryear{Skinner, Dolence, Burrows, Radice  \& Vartanyan}{Skinner et~al.}{2019}]{Skinner2019}
Skinner M.~A.,  Dolence J.~C.,  Burrows A.,  Radice D.,   Vartanyan D.,  2019, \mn@doi [The Astrophysical Journal Supplement Series] {10.3847/1538-4365/ab007f}, 241, 7

\bibitem[\protect\citeauthoryear{{Thompson}, {Quataert}  \& {Murray}}{{Thompson} et~al.}{2005}]{Thompson2005}
{Thompson} T.~A.,  {Quataert} E.,   {Murray} N.,  2005, \mn@doi [\apj] {10.1086/431923}, \href {https://ui.adsabs.harvard.edu/abs/2005ApJ...630..167T} {630, 167}

\bibitem[\protect\citeauthoryear{Wibking \& Krumholz}{Wibking \& Krumholz}{2022}]{Wibking2022}
Wibking B.~D.,  Krumholz M.~R.,  2022, \mn@doi [Monthly Notices of the Royal Astronomical Society] {10.1093/mnras/stac439}, 512, 1430

\bibitem[\protect\citeauthoryear{Zel'dovich \& Raizer}{Zel'dovich \& Raizer}{2002}]{Zeldovich2002}
Zel'dovich Y.~B.,  Raizer Y.~P.,  2002, Physics of {{Shock Waves}} and {{High-Temperature Hydrodynamic Phenomena}}.
{Courier Corporation}

\bibitem[\protect\citeauthoryear{Zel'dovich \& Raizer}{Zel'dovich \& Raizer}{2012}]{Zeldovich1967}
Zel'dovich Y.,  Raizer Y.,  2012, Physics of Shock Waves and High-Temperature Hydrodynamic Phenomena.
Dover Books on Physics, Dover Publications

\bibitem[\protect\citeauthoryear{Zhang}{Zhang}{2020}]{Zhang2020}
Zhang X.,  2020, \mn@doi [Research in Astronomy and Astrophysics] {10.1088/1674-4527/20/7/99}, 20, 099

\bibitem[\protect\citeauthoryear{{Zhang}, {Howell}, {Almgren}, {Burrows}  \& {Bell}}{{Zhang} et~al.}{2011}]{Zhang11a}
{Zhang} W.,  {Howell} L.,  {Almgren} A.,  {Burrows} A.,   {Bell} J.,  2011, \mn@doi [\apjs] {10.1088/0067-0049/196/2/20}, \href {http://adsabs.harvard.edu/abs/2011ApJS..196...20Z} {196, 20}

\makeatother
\end{thebibliography}

\appendix

\bsp	%
\label{lastpage}
\end{document}